\documentclass[%
superscriptaddress,
nofootinbib,
 amsmath,amssymb,
 aps,
fleqn
]{revtex4-2}

\usepackage{amssymb}
\usepackage{graphicx}% Include figure files
\usepackage{dcolumn}% Align table columns on decimal point
\usepackage{bm}% bold math
\usepackage[usenames,dvipsnames]{xcolor}
\usepackage[linkcolor=Blue,colorlinks=true, colorlinks=true, urlcolor=Blue, citecolor=Blue]{hyperref}
\usepackage[english]{babel}

\usepackage{tensor}
\usepackage[italicdiff]{physics}
\usepackage{amsthm}
\usepackage{subcaption}
\usepackage{cancel}

\usepackage{multirow}
\usepackage[disable]{todonotes}
\usepackage[]{mdframed}
\usepackage{xcolor}

\usepackage{todonotes}
\usepackage[skins, breakable]{tcolorbox}

\newcommand{\tb}[1]{\boldsymbol{#1}}

\makeatletter
\let\ORIbbl@fixname\bbl@fixname
\def\bbl@fixname#1{%
  \@ifundefined{languagealias@\expandafter\string#1}
    {\ORIbbl@fixname#1}
    {\edef\languagename{\@nameuse{languagealias@#1}}}%
}
\newcommand{\definelanguagealias}[2]{%
  \@namedef{languagealias@#1}{#2}%
}
\makeatother

\definelanguagealias{en}{english}

\newtheorem{mythm}{Theorem}[section]

\theoremstyle{definition}
\newtheorem{mydef}[mythm]{Definition}

\newcommand{\ph}{\mbox{\Huge\textvisiblespace}}
\newcommand{\pert}[2][1]{\delta^{(#1)}\!\!\left[#2\right]}
\newcommand{\pertnb}[1][1]{\delta^{(#1)}\!}
\numberwithin{equation}{section}

\newmdenv[linecolor = black, frametitle = \colorbox{white}{Waypoint}, frametitleaboveskip = -.7em, innertopmargin = -1.5em, shadow = true, shadowsize = 3pt, skipbelow = 15pt]{myEquationFrame}

\newmdenv[linecolor = black, frametitle = \colorbox{white}{Waypoint}, frametitleaboveskip = -.7em, innertopmargin = -.1em, shadow = true, shadowsize = 3pt, skipbelow = 15pt]{myTextFrame}

\begin{document}

\listoftodos

\preprint{APS/123-QED}

\title{Fundamental Cosmic Anisotropy and its Ramifications II \\ Perturbations in Bianchi spacetimes, and fixed in the Newtonian gauge}%

\author{Robbert W.\ Scholtens}
\email[contact author, ]{r.w.scholtens@rug.nl}
\affiliation{Kapteyn Astronomical Institute, University of Groningen, Groningen, The Netherlands}
\affiliation{Bernoulli Institute for Mathematics, Computer Science, and Artificial Intelligence, University of Groningen, Groningen, The Netherlands}

\author{Marcello Seri}
\email{m.seri@rug.nl}
\affiliation{Bernoulli Institute for Mathematics, Computer Science, and Artificial Intelligence, University of Groningen, Groningen, The Netherlands}

\author{Holger Waalkens}
\email{h.waalkens@rug.nl}
\affiliation{Bernoulli Institute for Mathematics, Computer Science, and Artificial Intelligence, University of Groningen, Groningen, The Netherlands}

\author{Rien van de Weygaert}
\email{weygaert@astro.rug.nl}
\affiliation{Kapteyn Astronomical Institute, University of Groningen, Groningen, The Netherlands}

\begin{abstract}
    \noindent The standard cosmological model is challenged by an ever-growing collection of observations, which invites (and stimulates) inquiry into possible additions and/or alterations. One such alteration comes from letting cosmic isotropy---as demanded by the cosmological principle---go, whilst maintaining only homogeneity. This study concerns Bianchi models, a class of anisotropic, homogeneous spacetimes, and in particular their perturbations. Knowledge of their properties under perturbations (such as allowed wavemodes) aids in understanding cosmological signatures of such universes, e.g.\ CMBs, and thus allows for comparsion to observation and the theory of the standard model. This study develops linear perturbation theory of general Bianchi models, by working in a frame such that metric components depend solely on (cosmic) time. Perturbation equations in the Newtonian gauge, but for arbitrary metric, are derived for energy density $\rho$, (relativistic) pressure $p$, momentum density $\tb{q}$, and anisotropic stress $\tb{\pi}$, for the case of scalar and pure tensor perturbations. For the former, the equations for density and pressure are combined to yield the equivalent of the Mukhanov-Sasaki equation for Bianchi models. For a specific choice of metric and fluid flow $\tb{u}$, the Friedmann equations for Bianchi models are also formulated, as this knowledge is necessary to fully formulate the perturbation equations. Finally, the obtained results are applied to the formulations of density contrasts in an Einstein-de Sitter universe and a Bianchi I universe.
\end{abstract}

% \abstract{Hello}

\maketitle
\tableofcontents

\section{Introduction}
Modern cosmology is based on the Cosmological Principle, stating that at the largest scales, the universe is spatially homogeneous and isotropic \cite[\S3]{peacock_cosmological_1999}, \cite{peebles_status_2024}, which heuristically means that every location in space is equivalent, as well as every spatial direction. From this, together with the Weyl principle \cite[\S3.4]{narlikar_introduction_2002} and Einstein's general relativity, one can then derive the \emph{Friedmann-Lemaître-Robertson-Walker} (FLRW) model---the backbone of the ``standard model of cosmology,'' $\Lambda$CDM. This model allows us to explain observations with great success, an example being given by the CMB power spectrum as measured by the \emph{Planck} mission \cite{planck_collaboration_planck_2020}.

Recently, though, certain challenges to this standard model have been emerging. These include an observed potential anisotropy in the Hubble parameter \cite{bolejko_differential_2016, migkas_probing_2020, boubel_testing_2025}, anisotropy in the cosmic acceleration \cite{colin_evidence_2019, rameez_anisotropy_2025}, anisotropies in quasar distributions \cite{secrest_test_2021, wagenveld_cosmic_2023} and supernovae \cite{sah_anisotropy_2024, colin_probing_2011}, and local bulk flows \cite{watkins_analysing_2023}. Peebles also gives an overview of ``anomalies'' not (yet) resolved by $\Lambda$CDM in \cite{peebles_anomalies_2022, peebles_status_2024}. There is also the excellent recent review by Kumar Aluri et al.\ on compatibility of the cosmological principle with observations \cite{kumar_aluri_is_2023}. As the sources cited above indicate, it may be that isotropy as postulated by the cosmological principle deserves some reconsideration, while still maintaining cosmic homogeneity---enter, thus, the homogeneous, \emph{anisotropic} cosmologies. Loosely speaking, these are cosmologies in which every spatial location is still equivalent, but spatial directions are no longer all equivalent; one could also say, there is (at least one) \emph{preferred} direction in space. These models can further be subdivided into types with varying degrees of rotational symmetry \cite[\S5.2]{ellis_cosmological_2008}.

In continuance of our previous work \cite{scholtens_fundamental_2025}, our primary interest is in the \emph{Bianchi} models, those generally without any additional symmetry other than homogeneity. Such models have been studied ever since their introduction in the 1960s \cite{ellis_class_1969, ellis_class_1970, maccallum_cosmological_1973,collins_why_1973,kundt_spatially_2003}---see also \cite{ellis_bianchi_2006} for a brief overview.

Our interests lay with \emph{perturbations} of such universes. Physically, cosmological perturbations manifest themselves as structures in the universe, such as the the CMB or the cosmic web, and observations of these structures forms the empirical basis for cosmological models. Understanding perturbations of Bianchi models, thus, leads to further understanding of the signatures of anisotropy. The presence of these signatures---or lack thereof---can then form a probe for assessing the validity of the Cosmological Principle, and to which extent (for instance, homogeneity scales).

%In particular, we are interested in \emph{perturbations} of such universes. Namely, routes to verifying whether or not (and then in how so far) the universe is isotropic at the largest scales lead to observing objects at the largest cosmological scales. Per definition these objects represent inhomogeneities, and as such a perturbative approach on top of any background framework is required.

In the FLRW model, this is a well-studied and well-understood field of inquiry---see for instance Peebles' seminal textbook \emph{The Large-Scale Structure of the Universe} \cite{peebles_large-scale_1980}. The study of perturbations of Bianchi models has mostly revolved around perturbations of the so-called Bianchi I model, which for distinct functions $a(t)$, $b(t)$, and $c(t)$, has a metric expressible as 
\begin{equation}
    \dd{s}^2=-\dd{t}^2+a^2(\dd{x})^2+b^2(\dd{y})^2+c^2(\dd{z})^2;
\end{equation}
see e.g. \cite{ludwick_gravitational_2024, le_delliou_anistropic_2020,hertzberg_constraints_2024,pereira_theory_2007}. These models are still quite tractable analytically, and allow some of the same analysis techniques as for FLRW. Another model of particular interest is Bianchi VII \cite{barrow_universal_1985, planck_collaboration_planck_2016, pontzen_rogues_2009}. Analyses of non-Bianchi I models usually assume some manner of closeness to FLRW, grounded in the current observations and closeness to isotropy, or single scale factor (despite, like above, multiple being allowed), to reduce complexity.

In this paper, we study perturbations of Bianchi spacetimes---not a priori restricted to a specific type---by working in an adapted frame (or tetrad) for our spacetime, and performing the analysis as is traditionally carried out for FLRW universes. Such frames have been considered at length in the literature \cite{taub_empty_1951, jantzen_dynamical_1979}, and provide a popular method of studying Bianchi spacetimes. In such a frame, one may then derive objects of interest such as the kinematical quantities and the Friedmann equations. More specifically, the frame we choose is such that
\begin{enumerate}
    \item the spacetime metric for our Bianchi model has metric components \emph{dependent solely on cosmic time $t$;} and
    \item the constituent vector fields form a Lie algebra.
\end{enumerate}
That this can be done is elaborated on, for instance, in our previous work \cite{scholtens_fundamental_2025}. With these two properties, the Einstein field equations reduce from partial to ordinary differential equations, improving solvability. The price that is being paid is a more complicated interpretation of the results, and that we need to take into account effects due to, in general, non-coordinate frame we are using. Nevertheless, in said frame we can then still proceed with a perturbation analysis, carried out in the usual fashion.

The main result of this work is \eqref{eq:friedmann-perts-newton}. For a perturbation in the Newtonian gauge,
\begin{equation}
    \tensor{\delta g}{_\mu_\nu}=-2(\phi+\psi)\,\tensor{u}{_\mu}\tensor{u}{_\nu}-2\psi\,\tensor{g}{_\mu_\nu},
\end{equation}
\eqref{eq:friedmann-perts-newton} is a set of perturbation equations for the relativistic energy density $\rho$, relativistic pressure $p$, momentum density $\tb{q}$, and anisotropic pressure $\tb{\pi}$ in terms of the perturbation and the kinematical quantities $\theta$, $\tb{\sigma}$, and $\tb{\omega}$. Under some further, mild assumptions---see \S\ref{sec:scalarPerts}---the perturbations of $\rho$ and $p$ may be combined into one equation that constrains the evolution of a perturbation:
\begin{equation}
    \begin{split}
        &\ddot{\psi}-c_s^2\lozenge\psi+\tfrac{1}{3}\tensor{\dot{u}}{^\alpha}\tensor{\psi}{_{;\alpha}}+\tfrac{4}{3}\theta\dot{\psi}+\left(\tfrac{4}{3}\dot{\theta}+(1+c_s^2)(\tfrac{2}{3}\theta^2-\Lambda)+2(1-c_s^2)\sigma^2+\tfrac{2}{3}(1-9c_s^2)\omega^2\right)\psi \\ 
        &\qquad=2\kappa(1-3c_s^2)\,\tensor{q}{_\alpha}\tensor{\pertnb u}{^\alpha}.
    \end{split}
\end{equation}
Here, $\lozenge$ is a Laplacian-like operator, defined in Definition \ref{def:pwo}. This equation generalizes the well-known Mukhanov-Sasaki equation for (conformal) FLRW spacetimes \cite[Eq.~(10.76)]{ellis_relativistic_2012}.

\subsection*{The Large-Scale Structure of this paper}
In \S\ref{sec:prelims}, we will elaborate on the main framework for our analysis (as briefly introduced above), namely that of geometry in non-coordinate frames and the complications this entails. Furthermore, we reintroduce the traditional 1+3 formalism of cosmology, introducing the kinematical tensors and the Raychaudhuri equation. Subsequently, in \S\ref{sec:derPerts}, we discuss briefly the mathematical context of (cosmic) perturbation theory, and then proceed with perturbing the metric, energy-momentum, and Einstein tensors. In these, we assume merely that the fluid flow $\tb{u}$ is normalized, $|\tb{u}|^2=-1$, and not its geodesy, or any properties of the kinematical tensors. Then, in \S\ref{sec:newton_gauge}, we fix our perturbations to the Newtonian gauge. We derive the perturbation equations for pure scalar and pure tensor perturbations, and through some mild assumptions combine the former into one equation, \eqref{eq:masterEquations}, governing scalar perturbations, akin to the Mukhanov-Sasaki perturbation equation. In \S\ref{sec:metricChoice}, we specialize to the case of a metric of purely diagonal form with distinct scale factors, $\tensor{g}{_\mu_\nu}=\operatorname{diag}(-1,a_1(t)^2,a_2(t)^2,a_3(t)^2)$, and stationary fluid flow, $\tensor{u}{^\mu}=\tensor*{\delta}{^\mu_0}$. We find the relevant kinematical tensors, as well as write down the Friedmann equations based on results by Wald \cite{wald_general_1984}. These Friedmann equations explicitly exhibit the non-coordinate nature of the frames, in the form of contributions based on the structure constants. We calculate both the kinematical tensors and the Friedmann equations explicitly here as they are required in the perturbation equations \eqref{eq:friedmann-perts-newton}. Finally, in \S\ref{sec:density_contrasts} we utilize the found results in order to provide some qualitative results for density contrasts in Einstein-de Sitter and Bianchi I universes. We end with a discussion and outlook to (our) future research.

\section{Preliminaries}\label{sec:prelims}

\subsection{Homogeneous geometries}\label{sec:homgeom}
Any spacetime metric $\tb{g}$ that is spatially homogeneous has---by definition---three spacelike Killing vector fields (KVFs) that at each event in spacetime are independent of each other. Under the assumption of simple transitivity,\footnote{That is, that when following the flow of one KVF, the flow of another KVF will not be intersected.} these generates a spacelike hypersurface in spacetime \cite[\S6.3]{ryan_homogeneous_1975}. Then, congruent with the assumption of global hyperbolicity, spacetime can \emph{globally} be split as $\Sigma^3\times\mathbb{R}$, i.e.\ ``$\text{spacetime}=\text{space}\times\text{time}$," or foliated with copies of that spacelike hypersurface. Note that this decomposition is purely \emph{topological;} for instance the precise embeddings of the $\mathbb{R}$ and $\Sigma^3$ into spacetime are not determined from this observation.

Using the aforementioned KVFs, it can be shown that the metric can always be written as
\begin{equation}\label{eq:sep-metric}
    \tb{g}(x)=\tensor{g}{_\alpha_\beta}(t)\,\tensor{\tb{e}}{^\alpha}\,\tensor{\tb{e}}{^\beta},\qq{with}\mathcal{L}_{(\tensor{\tb{e}}{^0})^*}\tensor{\tb{e}}{^\mu}=0,
\end{equation}
for suitable 1-forms $\{\tensor{\tb{e}}{^\mu}\}_{\mu=0}^3$, with $\tensor{\tb{e}}{^0}=\dd{t}$ exact---see e.g.\ our previous paper \cite{scholtens_cosmic_2024} for further details. We find a collection of vector fields $\{\tensor{\tb{X}}{_\mu}\}_{\mu=0}^3$ from the $\{\tensor{\tb{e}}{^\mu}\}$ by using the canonical isomorphism between tangent and cotangent spaces. The $\{\tensor{\tb{X}}{_\mu}\}$ form a frame at each event in spacetime, and moreover they form a Lie algebra, with the relations
\begin{equation}\label{eq:commRelations}
    [\tensor{\tb{X}}{_0},\tensor{\tb{X}}{_i}]=0\qand[\tensor{\tb{X}}{_i},\tensor{\tb{X}}{_j}]=\tensor{C}{^a_i_j}\tensor{\tb{X}}{_a},
\end{equation}
featuring the \emph{structure constants} $\tensor{C}{^k_i_j}$. (This follows if the methodology of \cite{scholtens_cosmic_2024} is utilized for finding the separated form \eqref{eq:sep-metric}.) 

The first relation of \eqref{eq:commRelations} implies that the flow of $\tensor{\tb{X}}{_0}$ can be used as a global coordinate cf.\ the 1-form $\tensor{\tb{e}}{^0}$ being exact, and that in contrast the relations among the vector fields $\{\tensor{\tb{X}}{_i}\}_{i=1}^3$ do not generally allow this coordinate-like interpretation. A frame in which not all of the structure constants vanish is thus called \emph{non-coordinated.} Physically, this means that the directions may be ``non-constant,'' by which we mean that their very definition---the vector field $\tensor{\tb{X}}{_i}$---may vary from place to place. But, by construction they are independent of time, so however they change, there is no temporal dependence.

The above formulation \eqref{eq:sep-metric} allows us to write any spatially homogeneous metric into a ``separated form,'' wherein the spatial information is absorbed into the basis of 1-forms, and the temporal information into the actual components of the metric. Clearly, we should prefer to work in this separated form over any other presented form of a spatially homogeneous metric: since the metric components depend only on time, this will reduce the Einstein equations from partial to ordinary differential equations---a huge simplification. The cost we pay is geometric non-triviality under the hood.
\begin{myTextFrame}[frametitle = \colorbox{white}{Non-coordinated frames}]
    We assume from now on that \emph{any frame is not necessarily coordinated,} but that it is has the properties outlined above. That is, that the frame for spacetime $\{\tensor{\tb{X}}{_\mu}\}_{\mu=0}^3$ satisfies:
    \begin{enumerate}
        \item $\tensor{\tb{X}}{_0}$ is timelike everywhere;
        \item $[\tensor{\tb{X}}{_0},\tensor{\tb{X}}{_\mu}]=\tb{0}$; and
        \item $\{\tensor{\tb{X}}{_i}\}_{i=1}^3$ forms a Lie algebra, with structure constants $[\tensor{\tb{X}}{_i},\tensor{\tb{X}}{_j}]=\tensor{C}{^a_i_j}\tensor{\tb{X}}{_a}$.
    \end{enumerate} 
\end{myTextFrame}
Note that coordinate frames also satisfy this definition, in which case the structure constants vanish identically (as the frame vector fields are partial derivatives of the coordinates, which commute).

One non-triviality comes from having to reconsider the Levi-Civita connection. On any (pseudo-)Riemannian manifold, it is the unique connection that is both compatible with metric and torsion-free. Its definition in a local frame is given by \cite[\S2.5]{ryan_homogeneous_1975}
\begin{equation}
    \tensor{\Gamma}{^\sigma_\mu_\nu}=\tfrac{1}{2}\tensor{g}{^\alpha^\sigma}\left(2\,\tensor{g}{_\alpha_{(\mu}_{,\nu)}}-\tensor{g}{_\mu_\nu_{,\alpha}}+2\,\tensor{C}{^\beta_\alpha_{(\mu}}\tensor{g}{_{\nu)}_\beta}\right)-\tfrac{1}{2}\tensor{C}{^\sigma_\mu_\nu},
\end{equation}
where $\tb{\Gamma}$ are the \emph{connection symbols} such that $\tensor{\nabla}{_\mu}\tensor{X}{_\nu}\equiv\tensor{\Gamma}{^\sigma_\nu_\mu}\tensor{X}{_\sigma}$. One can derive this result from e.g.\ Koszul's formula---see \cite[\S5]{lee_introduction_2010}. This reduces the known formula for the Christoffel symbols in the case the frame is coordinate. Note that the structure constants of our frame are introduced as proportional to the metric, instead of a derivative, and also a constant term is added. This destroys the symmetry in the lower two indices of the Levi-Civita connection; or, the Levi-Civita connection is symmetric if and only if all structure constants vanish, i.e.\ that our frame is coordinated.

The definition of the Riemann tensor is also changed,
\begin{equation}\label{eq:newRiemann}
    \tensor{R}{^\sigma_\rho_\mu_\nu}=\tensor{\Gamma}{^\sigma_\rho_\nu_{,\mu}}-\tensor{\Gamma}{^\sigma_\rho_\mu_{,\nu}}+\tensor{\Gamma}{^\alpha_\rho_\nu}\tensor{\Gamma}{^\sigma_\alpha_\mu}-\tensor{\Gamma}{^\alpha_\rho_\mu}\tensor{\Gamma}{^\sigma_\alpha_\nu}-\tensor{C}{^\alpha_\mu_\nu}\tensor{\Gamma}{^\sigma_\rho_\alpha},
\end{equation}
wherein the non-coordinated nature of our basis is reflected explicitly in the final term (and implicitly in the definition of the Christoffel symbols). The above formula may be derived from the general formula for the elements of the Riemann tensor,
\begin{equation}
    \tensor{R}{_\sigma_\rho_\mu_\nu}:=\tb{g}(\tb{R}(\tensor{\tb{X}}{_\mu},\tensor{\tb{X}}{_\nu})\tensor{\tb{X}}{_\rho},\tensor{\tb{X}}{_\sigma}),\qq{where}\tb{R}(\tb{Y},\tb{Z})=\nabla_{\tb{Y}}\nabla_{\tb{Z}}-\nabla_{\tb{Z}}\nabla_{\tb{Y}}-\nabla_{[\tb{Y},\tb{Z}]},
\end{equation}
which holds also for non-coordinated frames. Importantly, by direct verification we see that the \emph{Ricci identity for the Riemann tensor,}
\begin{equation}\label{eq:ricciIdentity}
    2\,\tensor{u}{^\sigma_{;[\mu\nu]}}=\tensor{R}{^\sigma_\alpha_\nu_\mu}\tensor{u}{^\alpha},
\end{equation}
also remains valid in non-coordinated frames. This will be of importance below in \S\ref{sec:kinematicalTensors}.

\subsection{The 1+3 decomposition}\label{sec:3+1}
Not only is a cosmological model we choose to study dependent on its metric, but also on the inherent movement of so-called ``fundamental particles'': large-scale objects such as galaxy clusters which are assumed to move under gravity alone. In this way, the cosmos can be seen as a ``fluid'' with these fundamental particles as the ``molecules'' that make it up. (A more refined version of this statement is known as Weyl's postulate \cite[\S3.5.1]{narlikar_introduction_2002}.) The 4-velocity of this fluid flow mathematically comes in the form of a vector field on spacetime, and is denoted by $\tb{u}$. We can also view $\tb{u}$ as a family of observers, in which case its flow can be seen as an observer's world line with tangent vector $\tb{u}$ along it \cite[\S4.2]{ellis_relativistic_2012}.

The choice of fluid flow, being related to an observer's worldline, influences among others the notion of energy density $\rho$ and pressure $p$ appearing in the Friedmann equations of the cosmological model. This can be seen, for instance, by examining the energy-momentum tensor for a perfect fluid,
\begin{equation}
    \tensor{T}{_\mu_\nu}=(\rho+p)\,\tensor{u}{_\mu}\tensor{u}{_\nu}+p\,\tensor{g}{_\mu_\nu}.
\end{equation}
Hence, different choices of $\tb{u}$ will eventually lead to different values for $\rho$ and $p$, when writing down the Friedmann equations.

Though in principle any choice of $\tb{u}$ can be made from a mathematical point of view, a physical requirement is that the fluid flow is timelike, i.e.\ that $\tensor{u}{^\alpha}\tensor{u}{_\alpha}<0$ at every event---if it were not, it could not be tangent to observers' worldlines. In fact, we shall normalize the fluid flow (which is a matter of reparametrization), so that we make the common assumption that at every event, $\tensor{u}{^\alpha}\tensor{u}{_\alpha}=-1$.

Another common assumption is the one of geodesic fluid flow, i.e.\ fluid flows wherein the condition $\tensor{\dot{u}}{^\mu}:=\tensor{u}{^\alpha}\tensor{u}{^\mu_{;\alpha}}\equiv0$. This corresponds to observers along the fluid flow to be in constant free fall, i.e.\ free movement along geodesics in spacetime. However, in this work we elect \emph{not} to make this assumption, and instead carry $\tensor{\dot{u}}{^\mu}$ through the calculations. If at a later stage we wish to consider geodesic fluid flows instead, we may then at that moment set $\tensor{\dot{u}}{^\mu}\equiv0$.

In the field of cosmology, often the particular choice of fluid flow is made as $\tensor{u}{^\mu}=\tensor*{\delta}{^\mu_0}=(1,0,0,0)$, for $\tensor{x}{^0}=t$ the cosmic time. This corresponds to the cosmic fluid being at rest on the largest scales. The rationale is that for the timelike object with 4-velocity $\tb{u}$, a frame featuring $\tb{u}$ as one of the basis vector fields can be defined; in that frame, then, the object is spatially stationary. Thus, the object only has motion in the timelike direction, which can subsequently be normalized (e.g.\ corresponding to a choice of metric coefficient $\tensor{g}{_0_0}$). This also synergizes well with the cosmological principle, which through isotropy forbids any preferred spatial directions. If we did choose a frame in which $\tb{u}$ corresponded to motion through space, the projection of $\tb{u}$ onto space could be argued to be a preferred spatial direction, contradicting isotropy. Since in our studies we are interested especially in letting go of the spatial isotropy assumption, we do \emph{not} assume that $\tensor{u}{^\mu}\propto\tensor*{\delta}{^\mu_0}$ and only assume normalization.

Now that we have established that there is a preferred timelike direction at every event, being the fluid flow $\tb{u}$, it makes intuitive and mathematical sense to split tensorial quantities according to their components along and perpendicular to this fluid flow. This is what we shall develop for the rest of this section.

\subsubsection{Decomposing tensors}
In order to study the components of a tensor along and perpendicular to the fluid flow $\tb{u}$, we need an operator that facilitates this splitting. For this we introduce the \emph{projection operator} $\tb{h}$, as
\begin{equation}\label{eq:projection_op}
    \tensor{h}{_\mu_\nu}:=\tensor{g}{_\mu_\nu}+\tensor{u}{_\mu}\tensor{u}{_\nu},
\end{equation}
which has the properties that
\begin{equation}
    \tensor{h}{_{[\mu\nu]}}=0,\quad\tensor{u}{^\alpha}\tensor*{h}{^\mu_\alpha}=0,\qand\tensor*{h}{^\mu_\alpha}\tensor*{h}{^\alpha_\nu}=\tensor*{h}{^\mu_\nu}
\end{equation}
so that indeed $\tb{h}$ projects a vector onto a subspace perpendicular to $\tb{u}$ in the tangent space above each event in spacetime. In fact, since $\tb{h}$ can be viewed as the pullback of $\tb{g}$ under the embedding of this subspace into spacetime \cite[\S2.9]{ellis_relativistic_2012}, it follows that $\tb{h}$ projects vectors onto a hypersurface. Note that $\tb{h}$ is the first fundamental form of the hypersurface. The projection tensor $\tb{h}$ can be used as the metric on this hypersurface, and, since the sole timelike direction has been projected away, this metric is Riemannian.

Let us now do some projections. Assume that we wish to project the symmetric 2-tensor $\tensor{\mathcal{T}}{_\mu_\nu}$, which we record has ten degrees of freedom. The projection decomposition is found through calculating
\begin{equation}
    \begin{split}
        \tensor{\mathcal{T}}{_\mu_\nu}=\tensor*{\delta}{^\alpha_\mu}\tensor*{\delta}{^\beta_\nu}\tensor{\mathcal{T}}{_\alpha_\beta}&=(\tensor*{h}{^\alpha_\mu}-\tensor{u}{^\alpha}\tensor{u}{_\mu})(\tensor*{h}{^\beta_\nu}-\tensor{u}{^\beta}\tensor{u}{_\nu})\tensor{\mathcal{T}}{_\alpha_\beta} \\
        &=\tensor{u}{^\alpha}\tensor{u}{^\beta}\tensor{\mathcal{T}}{_\alpha_\beta}\tensor{u}{_\mu}\tensor{u}{_\nu}-2\,\tensor{u}{^\alpha}\tensor*{h}{^\beta_\gamma}\tensor{\mathcal{T}}{_\alpha_\beta}\tensor{u}{_{(\mu}}\tensor*{h}{^\gamma_{\nu)}}+\tensor*{h}{^\alpha_\gamma}\tensor*{h}{^\beta_\delta}\tensor{\mathcal{T}}{_\alpha_\beta}\tensor*{h}{^\gamma_{(\mu}}\tensor*{h}{^\delta_{\nu)}}.
    \end{split}
\end{equation}
In matrix notation, this corresponds to
\begin{equation}
    \tensor{\mathcal{T}}{_\mu_\nu}=\left(\begin{array}{c|ccc} \tensor{u}{^\alpha}\tensor{u}{^\beta}\tensor{\mathcal{T}}{_\alpha_\beta} & \cdots & \tensor{u}{^\alpha}\tensor*{h}{^\beta_\rho}\tensor{\mathcal{T}}{_\alpha_\beta} & \cdots \\ \hline \vdots & & & \\ \tensor{u}{^\alpha}\tensor*{h}{^\beta^\rho}\tensor{\mathcal{T}}{_\alpha_\beta} & & \tensor*{h}{^\alpha_\sigma}\tensor*{h}{^\beta_\tau}\tensor{\mathcal{T}}{_\alpha_\beta} & \\ \vdots & & & \end{array}\right),
\end{equation}
with the usual word of warning that this should only be viewed as an array of quantities, and not as an operator on vectors. (For that to be the case, the valence needs to be $(1,1)$.) However, for interpretation's sake it will be convenient to rewrite this expression into
\begin{equation}\label{eq:tensorDecomp}
    \tensor{\mathcal{T}}{_\mu_\nu}=\tensor{u}{^\alpha}\tensor{u}{^\beta}\tensor{\mathcal{T}}{_\alpha_\beta}\tensor{u}{_\mu}\tensor{u}{_\nu}-2\,\tensor{u}{^\alpha}\tensor*{h}{^\beta_\epsilon}\tensor{\mathcal{T}}{_\alpha_\beta}\tensor{u}{_{(\mu}}\tensor*{h}{^\epsilon_{\nu)}}+\tfrac{1}{3}\tensor{h}{^\alpha^\beta}\tensor{\mathcal{T}}{_\alpha_\beta}\tensor{h}{_\mu_\nu}+(\tensor*{h}{^\alpha_{(\mu}}\tensor*{h}{^\beta_{\nu)}}-\tfrac{1}{3}\tensor{h}{^\alpha^\beta}\tensor{h}{_\mu_\nu})\tensor{\mathcal{T}}{_\alpha_\beta}.
\end{equation}
The term $\tfrac{1}{3}\tensor{h}{^\alpha^\beta}\tensor{\mathcal{T}}{_\alpha_\beta}$ is the trace of $\mathcal{T}$'s projection onto the hypersurface, which quantifies the ``size'' of the tensor $\mathcal{T}$ on the hypersurface. In introducing this term into the decomposition, we inadvertently created the final term in the decomposition, $(\tensor*{h}{^\alpha_{(\mu}}\tensor*{h}{^\beta_{\nu)}}-\tfrac{1}{3}\tensor{h}{^\alpha^\beta}\tensor{h}{_\mu_\nu})\tensor{\mathcal{T}}{_\alpha_\beta}$. By inspection we note that this term lives on the perpendicular subspace, is symmetric,\footnote{This observation is superfluous in our case since $\mathcal{T}$ was symmetric to begin with.} and also traceless. Since this operator pops up often enough in the context of 1+3 decompositions, it has its own notation and name: the \emph{projected, symmetric, tracefree} (PSTF) part. It is defined for a (not necessarily symmetric) rank 2 tensor by precisely the final summand in \eqref{eq:tensorDecomp}:
\begin{equation}
    \tensor{\mathcal{T}}{_{\langle\mu\nu\rangle}}:=(\tensor*{h}{^\alpha_{(\mu}}\tensor*{h}{^\beta_{\nu)}}-\tfrac{1}{3}\tensor{h}{^\alpha^\beta}\tensor{h}{_\mu_\nu})\tensor{\mathcal{T}}{_\alpha_\beta}.
\end{equation}
Hence, we can present the final form of symmetric tensor decomposition as
\begin{equation}
    \tensor{\mathcal{T}}{_\mu_\nu}=\tensor{u}{^\alpha}\tensor{u}{^\beta}\tensor{\mathcal{T}}{_\alpha_\beta}\tensor{u}{_\mu}\tensor{u}{_\nu}+\tfrac{1}{3}\tensor{h}{^\alpha^\beta}\tensor{\mathcal{T}}{_\alpha_\beta}\tensor{h}{_\mu_\nu}-2\,\tensor{u}{^\alpha}\tensor*{h}{^\beta_\epsilon}\tensor{\mathcal{T}}{_\alpha_\beta}\tensor{u}{_{(\mu}}\tensor*{h}{^\epsilon_{\nu)}}+\tensor{\mathcal{T}}{_{\langle\mu\nu\rangle}}.
\end{equation}

\subsubsection{The wave operator and its projection(s)}
We shall also be interested in the projection properties of derivatives and differential operators. Naturally, these will tell us about the projected changes in quantities of interest. Given standard theory, the \emph{wave operator} or \emph{d'Alembertian} is defined by the formula
\begin{equation}
    \square\,f:=\tensor{g}{^\alpha^\beta}\,\tensor{\nabla}{_\alpha}\!\tensor{\nabla}{_\beta}f,
\end{equation}
for scalar functions $f$, and analogously for tensors. It is so called since in the Minkowski metric, the kernel of $\square$ consists of precisely those functions which satisfy the Euclidean wave equation. It shall turn out that we want to consider a decomposition of this operator utilizing the fluid flow that we have defined, thereby obtaining an operator ``along'' the fluid flow, and an operator for the perpendicular hypersurfaces. The most straightforward way to carry this out is by decomposing the metric $\tb{g}$ in the above definition of the wave operator:
\begin{equation}
    \square\,f=\tensor{h}{^\alpha^\beta}\,\tensor{\nabla}{_\alpha}\!\tensor{\nabla}{_\beta}f-\tensor{u}{^\alpha}\tensor{u}{^\beta}\,\tensor{\nabla}{_\alpha}\!\tensor{\nabla}{_\beta}f=\tensor{h}{^\alpha^\beta}\,\tensor{\nabla}{_\alpha}\!\tensor{\nabla}{_\beta}f-\ddot{f}+\tensor{\dot{u}}{^\alpha}\tensor{f}{_{;\alpha}}.
\end{equation}
(Here and from now on, the dot operator on any tensor indicates differentiation along the fluid flow.) Since the hypersurface onto which we project with $\tb{h}$ is spatial (any tangent vector to it is spacelike), we \emph{could} conclude that $\tensor{h}{^\alpha^\beta}\,\tensor{\nabla}{_\alpha}\!\tensor{\nabla}{_\beta}$ is the Laplace-Beltrami operator (or Laplacian) for the hypersurface.

This would go contrary to the literature, though. The definition of the Laplacian of a scalar function $f$ is $\triangle f:=\operatorname{div}(\operatorname{grad} f)$ \cite[\S2.1]{petersen_riemannian_2006}. If one thus wants to consider the Laplacian restricted to a submanifold (such as our perpendicular hypersurface is, through the embedding into spacetime), the divergence and gradient should be taken on the hypersurface. That is, by using a projected covariant derivative: $\tensor{D}{_\mu}f:=\tensor*{h}{^\alpha_\mu}\,\tensor{\nabla}{_\alpha}f$ for scalar functions $f$. For tensorial objects, the definition becomes slightly more complicated \cite[\S4.4.2]{ellis_relativistic_2012}:
\begin{equation}
    \tensor{D}{_\rho}\tensor{\mathcal{T}}{^\mu^\dots_\nu_\dots}=\tensor*{h}{^\alpha_\nu}\tensor*{h}{^\mu_\beta}\tensor*{h}{_\rho^\gamma}\cdots\tensor{\nabla}{_\gamma}\tensor{\mathcal{T}}{^\beta^\dots_\alpha_\dots},
\end{equation}
i.e.\ \emph{each index} is projected after application of $D$, not only the index of $D$. A common shorthand, akin to the semicolon notation, is to use a subscript $|$ in order to indicate this derivative: $\ph_{|\mu}\equiv\tensor{D}{_\mu}\ph$. Keeping line with the original definition, then, the Laplacian should be defined as $\tensor{D}{^\alpha}\tensor{D}{_\alpha}$.

To recall further below and name these two concepts, we have the following Definition.
\begin{mydef}[Wave operators on the perpendicular hypersurface]\label{def:pwo}
    For a scalar function $f$, we define
    \begin{enumerate}
        \item $\triangle f:=\tensor{h}{^\alpha^\beta}\tensor{D}{_\alpha}\tensor{D}{_\beta}f$ as the \emph{Laplacian} on the hypersurface perpendicular to the fluid flow; and
        \item $\lozenge f:=\tensor{h}{^\alpha^\beta}\,\tensor{\nabla}{_\alpha}\!\tensor{\nabla}{_\beta}f$ as the \emph{projected wave operator.}
    \end{enumerate}
    These definitions also hold for general tensorial objects.
\end{mydef}
We also note the relation between the two operators for scalar functions: $\triangle f=\lozenge f+\theta\dot{f}$.

\subsubsection{The kinematical tensors}\label{sec:kinematicalTensors}
One particularly important object to study is the covariant derivative of the fluid flow, and its projection. as this gives us insight in the relative movement of nearby fundamental particles---see \cite[\S2.4]{ellis_republication_2009}. We proceed for $\tensor{u}{_\mu_{;\nu}}$ in the same way as we did for the projections above:
\begin{equation}
    \tensor{u}{_\mu_{;\nu}}=(\tensor*{h}{^\alpha_\mu}-\tensor{u}{^\alpha}\tensor{u}{_\mu})(\tensor*{h}{^\beta_\nu}-\tensor{u}{^\beta}\tensor{u}{_\nu})\tensor{u}{_\alpha_{;\beta}}=\tensor{u}{_\mu_{|\nu}}+\tensor{u}{^\beta}\tensor*{h}{^\alpha_\mu_{;\beta}}\tensor{u}{_\alpha}\tensor{u}{_\nu}=\tensor{u}{_\mu_{|\nu}}-\tensor{\dot{u}}{_\mu}\tensor{u}{_\nu}.
\end{equation}
Remarkably, there is no $\tb{u}\tb{u}$-component to $\tensor{u}{_\mu_{;\nu}}$, for $\tensor{u}{_\alpha}\tensor{\dot{u}}{^\alpha}=\tfrac{1}{2}\tensor{u}{^\beta}\tensor{\nabla}{_\beta}(\tensor{u}{_\alpha}\tensor{u}{^\alpha})=0$. We can then proceed with further decomposing the projected fluid flow $\tensor{u}{_\mu_{|\nu}}$, namely to the form of
\begin{equation}\label{eq:u-decomp}
    \tensor{u}{_\mu_{|\nu}}=\tfrac{1}{3}\theta\,\tensor{h}{_\mu_\nu}+\tensor{\sigma}{_\mu_\nu}+\tensor{\omega}{_\mu_\nu},
\end{equation}
where (i) $\theta:=\tensor{u}{^\alpha_{|\alpha}}$ is the \emph{expansion scalar,} (ii) $\tensor{\sigma}{_\mu_\nu}:=\tensor{u}{_{\langle\mu|\nu\rangle}}$ is the \emph{shear tensor,} and (iii) $\tensor{\omega}{_\mu_\nu}=\tensor{u}{_{[\mu|\nu]}}$ is the \emph{vorticity tensor.} From the shear and vorticity tensors, we can create the additional scalars
\begin{equation}
    \sigma^2:=\tfrac{1}{2}\tensor{\sigma}{^\alpha^\beta}\tensor{\sigma}{_\alpha_\beta}\qand\omega^2=\tfrac{1}{2}\tensor{\omega}{^\alpha^\beta}\tensor{\omega}{_\alpha_\beta}.
\end{equation}

Physical interpretations of these quantities are discussed in e.g.\ \cite[\S2.5]{ellis_republication_2009}. In summary, (i) the expansion scalar $\theta$ quantifies the expansion/contraction of a fluid element as it travels along its worldline, (ii) the shear $\tb{\sigma}$ describes the change in its shape normalized to constant volume, and (iii) the vorticity $\tb{\omega}$ indicates its rotation. 
% A visual illustration is given in Figure \ref{fig:kinematicalEffects}.
% \begin{figure}[t]
%     \centering
%     \includegraphics[width = .5\textwidth]{pictures/kinematicalEffects.png}
%     \caption{Interpretations of the kinematical tensors, from \cite{ellis_republication_2009}. The original caption reads: ``During a small time interval, a) the action of $\theta$ alone transforms a fluid sphere to a similar sphere of different volume but with the same orientation. b) the action of $\tensor{\sigma}{_\mu_\nu}$ alone distorts the sphere, leavings its volume constant and the directions of the principal axes of shear unchanged. c) The action of $\tensor{\omega}{_\mu_\nu}$ [ed.] alone is to give rigid rotation leaving one direction (the axis of rotation) fixed.''}
%     \label{fig:kinematicalEffects}
% \end{figure}

This decomposition moreover paves the way for the Raychaudhuri equation. This couples the kinematical scalars, differentiated fluid flow, and the Ricci tensor contracted along the fluid direction. The traditional way to derive it is to contract the Ricci identity for the Riemann tensor \eqref{eq:ricciIdentity}: first contracting the Riemann tensor, and then contractin with another $\tb{u}$. The result is the following:
\begin{myEquationFrame}[frametitle = \colorbox{white}{Raychaudhuri equation}]
    \begin{equation}
        \dot{\theta}+\tensor{u}{^\alpha_{|\beta}}\tensor{u}{^\beta_{|\alpha}}-\tensor{\dot{u}}{^\alpha_{;\alpha}}+\tensor{R}{_\alpha_\beta}\tensor{u}{^\alpha}\tensor{u}{^\beta}=0=\dot{\theta}+\tfrac{1}{3}\theta^2+2(\sigma^2-\omega^2)-\tensor{\dot{u}}{^\alpha_{;\alpha}}+\tensor{R}{_\alpha_\beta}\tensor{u}{^\alpha}\tensor{u}{^\beta},
    \end{equation}
\end{myEquationFrame}
where $\tensor{R}{_\mu_\nu}$ indicates the Ricci tensor. Additionally, one can use the Einstein equation in order to substitute for the contracted Ricci tensor quantities related to the matter content, e.g.\ as expanded on in \cite[\S6.1]{ellis_relativistic_2012}. We emphasize that this equation holds regardless of the frame that is being used, due to the Ricci identity still holding. The non-coordinate nature of a frame is exhibited in this equation through the frequent use of covariant differentiation.

% To relate these two notions, we can make the following calculation:
% \begin{equation}
%     \tensor{h}{^\mu^\nu}\tensor{D}{_\mu}\tensor{D}{_\nu}=\tensor{h}{^\mu^\nu}\,\tensor{\nabla}{_\mu}\!\tensor{\nabla}{_\nu}+\tensor{h}{^\alpha^\beta}\tensor{u}{_{\alpha;\beta}}\,\tensor{u}{^\mu}\,\tensor{\nabla}{_\mu}=\tensor{h}{^\mu^\nu}\,\tensor{\nabla}{_\mu}\!\tensor{\nabla}{_\nu}+\theta\,\tensor{u}{^\mu}\,\tensor{\nabla}{_\mu}
% \end{equation}

% This decomposition paves the way for the \emph{Raychaudhuri equation}, an equation most important and insightful. The traditional way of deriving this equation is through contraction of the Ricci identity,
% \begin{equation}
%     \tensor{u}{^\alpha_{;[\beta\gamma]}}=\tensor{R}{^\alpha_\delta_\beta_\gamma}\tensor{u}{^\delta}
% \end{equation}
% with the help of the given decomposition. However, since the Ricci identity is not precisely at our disposal (it only works when the metric frame is coordinatized), we cannot reproduce the Raychaudhuri equation this way. In a recent work \cite{agashe_lagrangian_2023}, the Raychaudhuri equation was found for such frames. Adapting their equation to our setting, we find that the Raychaudhuri equation for us is given by
% \begin{equation}
%     \dot{\theta}+\tfrac{1}{3}\theta^2+2\sigma^2-2\omega^2-2\tensor{C}{^\alpha_\beta_\gamma}\tensor{u}{^\beta}\tensor{u}{^\gamma_{;\alpha}}+\tensor{R}{_\alpha_\beta}\tensor{u}{^\alpha}\tensor{u}{^\beta}=0.
% \end{equation}

\section{Deriving perturbations}\label{sec:derPerts}
The object of cosmic perturbation theory is to formulate and solve the perturbed Einstein equations,
\begin{equation}\label{eq:einstein_eq_pert}
    \tensor{G}{_\mu_\nu}+\Lambda\,\tensor{g}{_\mu_\nu}=\kappa\,\tensor{T}{_\mu_\nu}\longrightarrow\tensor{\delta G}{_\mu_\nu}+\Lambda\,\tensor{\delta g}{_\mu_\nu}=\kappa\,\tensor{\delta T}{_\mu_\nu},
\end{equation}
under a perturbation of the metric $\tensor{g}{_\mu_\nu}\to\tensor{g}{_\mu_\nu}+\tensor{\delta g}{_\mu_\nu}$ \cite{mukhanov_theory_1992}, where $\kappa$ is a proportionality constant (in the $(0,2)$-valence, it takes the value $\kappa:=8\pi G/c^4$). As many have remarked: though the idea is simple, the hurdle is mostly computational.

We shall first briefly discuss the mathematical context in which (cosmological) perturbation theory is placed, letting us define perturbations in a solid mathematical manner. Subsequently, we calculate and simplify the perturbations (to first order) in the (i) spacetime metric, (ii) energy-momentum tensor (EMT), and (iii) the Einstein tensor, for which we also use an instance of the Mathematica package \texttt{xPand} \cite{pitrou_xpand_2013}. We can then formulate the perturbed Einstein equations \eqref{eq:einstein_eq_pert}, which are then to be solved in the subsequent section where we fix the Newtonian gauge.

\subsection{Perturbation theory, briefly}\label{sec:tensorPerts}
For this brief introduction, we follow the paper by Bruni et al.\ \cite{bruni_perturbations_1997}, although many more sources on this topic exist. One should view a perturbed spacetime as an element of a one-parameter family of spacetimes; let these spacetimes be $(\tensor[^{(\epsilon)}]{\mathcal{M}}{},\tensor[^{(\epsilon)}]{\tb{\tau}}{})$, for $\tb{\tau}$ some tensor of interest (or representing a set of tensors). Furthermore, $\epsilon$ is the relevant order parameter, $\epsilon\equiv0$ is the background value of the tensor $\tb{\tau}$, and $\tensor[^{(0)}]{\mathcal{M}}{}\cong\tensor[^{(\epsilon)}]{\mathcal{M}}{}$ is assumed for all $\epsilon$. Owing to this last relation, it is natural to consider
\begin{equation}
    \mathcal{N}=\bigsqcup_{\epsilon\in\mathbb{R}}\tensor[^{(\epsilon)}]{\mathcal{M}}{}\times\{\epsilon\}\cong\tensor[^{(0)}]{\mathcal{M}}{}\times\mathbb{R}
\end{equation}
as the collection of ``all'' the spacetimes of this perturbation family, indexed by $\epsilon$ (on the added real axis). The natural differentiable structure on $\mathcal{N}$ is the direct product of those on $\tensor[^{(0)}]{\mathcal{M}}{}$ and $\mathbb{R}$. Naturally, we can define $\tb{\tau}(x,\epsilon):=\tensor[^{(\epsilon)}]{\tb{\tau}}{}(x)$ on $\mathcal{N}$ utilizing our family of tensors.

Although we refer in this context to spacetimes with $\epsilon\neq0$ as ``perturbed,'' perturbations only make sense in comparison to situations that are (by definition) unperturbed, and we have not defined these comparisons mathematically yet. Intuitively, the comparisons we wish to study are w.r.t.\ to the background, $\epsilon=0$; thus, the perturbation in $\tensor[^{(0)}]{\tb{\tau}}{}$ is the difference ``$\tensor[^{(\epsilon)}]{\tb{\tau}}{}-\tensor[^{(0)}]{\tb{\tau}}{}$,'' and so that is the object that we wish to study. Since the two objects at hand live in different leaves of the foliation $\mathcal{N}$, however, direct comparison of the form above in quotations is not defined. Instead, to effect comparison, we need to pull the perturbed spacetime back to the background, where we can then look at the difference with the original spacetime. To this end, introduce a diffeomorphism $\alpha_\epsilon:\mathcal{N}\to\mathcal{N}$ so that $\alpha_\epsilon|_{\tensor[^{(0)}]{\mathcal{M}}{}}:\tensor[^{(0)}]{\mathcal{M}}{}\to\tensor[^{(\epsilon)}]{\mathcal{M}}{}$. Then the comparison is given by
\begin{equation}
    \Delta\tb{\tau}_\epsilon:=\alpha_\epsilon^*\tb{\tau}|_{\tensor[^{(0)}]{\mathcal{M}}{}}-\tensor[^{(0)}]{\tb{\tau}}{}
\end{equation}
thus defining the perturbation $\Delta\tb{\tau}_\epsilon$. So defined, $\Delta\tb{\tau}_\epsilon$ is a tensor on the background $\tensor[^{(0)}]{\mathcal{M}}{}$. Making the assumption that $\alpha$ is generated by some vector field $\tb{X}$ \emph{on $\mathcal{N}$, which is everywhere transverse to the leaves,} Bruni et al.\ show that this pullback can be written as
\begin{equation}
    \alpha_\epsilon^*\tb{\tau}|_{\tensor[^{(0)}]{\mathcal{M}}{}}=\sum_{k=0}^\infty\frac{\epsilon^k}{k!}\dv[i]{\alpha^*_\epsilon\tb{\tau}}{\epsilon}\Big|_{\epsilon=0,\tensor[^{(0)}]{\mathcal{M}}{}}=\sum_{k=0}^\infty\frac{\epsilon^k}{k!}\left[\mathcal{L}^k_X\tb{\tau}\right]_{\tensor[^{(0)}]{\mathcal{M}}{}},
\end{equation}
where $\mathcal{L}^k_X$ is the Lie derivative w.r.t.\ $\tb{X}$, applied $k$ times. Then we can make the definitions $\pert[k]{\tb{\tau}}:=\left[\mathcal{L}^k_{\tb{X}}\tb{\tau}\right]_{\tensor[^{(0)}]{\mathcal{M}}{}}$ to represent the $k$th order perturbation to $\tb{\tau}$, and we obtain the series
\begin{equation}\label{eq:pertExpansion}
    \alpha_\epsilon^*\tb{\tau}|_{^{(0)}\!\mathcal{M}}=\,^{(0)}\tb{\tau}+\epsilon\,\pert[1]{\tb{\tau}}+\tfrac{1}{2}\epsilon^2\,\pert[2]{\tb{\tau}}+\order{\epsilon^3}
\end{equation}
This expansion defines the mathematical context in which the perturbation to $\tb{\tau}$, an arbitrary tensor of interest, is defined. We now proceed with applying this expansion to the metric tensor, and subsequently the energy-momentum and Einstein tensors.

\subsection{The metric tensor}
Let us assume that the full metric of spacetime, $\tb{g}$, is decomposed in terms of a background metric $\overline{\tb{g}}$ and some perturbative expansion. That is, we write
\begin{equation}
    \tensor{g}{_\mu_\nu}=\tensor{\overline{g}}{_\mu_\nu}+\epsilon\,\pert{\tensor{g}{_\mu_\nu}}+\order{\epsilon^2},
\end{equation}
where $\epsilon$ is a formal expansion parameter. We formally include perturbations of higher order, although we shall limit ourselves only to perturbations in the first order. We can then do a similar expansion for the inverse metric, and demand that even in the perturbed setting the inverse relation holds:
\begin{equation}\label{eq:metricInvRelation}
    \tensor{g}{^\mu^\nu}=\tensor{\overline{g}}{^\mu^\nu}+\pert{\tensor{g}{^\mu^\nu}}+\order{\epsilon^2};\quad\tensor{g}{^\mu^\alpha}\tensor{g}{_\alpha_\nu}\stackrel{!}{=}\tensor*{\delta}{^\mu_\nu}\implies\begin{cases}\order{\epsilon^0}:&\tensor{\overline{g}}{^\mu^\alpha}\tensor{\overline{g}}{_\alpha_\nu}\stackrel{!}{=}\tensor*{\delta}{^\mu_\nu} \\ \order{\epsilon^1}:& \pert{\tensor{g}{^\mu^\alpha}}\,\tensor{\overline{g}}{_\alpha_\nu}+\tensor{\overline{g}}{^\mu^\alpha}\,\pert{\tensor{g}{_\alpha_\nu}}\stackrel{!}{=}0 \\ & \vdots\end{cases}
\end{equation}
The dots indicate that in order to effect perfect equality, one would need to solve the infinitude of equations for all orders of $\epsilon$ that follow. Truncating the solution to, for our purposes, first order implies that the relation of inverses holds only up to second order.

The zeroth order equation is satisfied by virtue of $\overline{\tb{g}}$ being a background metric. The first order equation can be solved for either the $(0,2)$- or $(2,0)$-valence, first order perturbation in terms of the other. Since the metric with $(0,2)$-valence is more fundamental (reflecting its definition as acting on the tangent bundle), we shall solve the second relation in \eqref{eq:metricInvRelation} for the upper index configuration in terms of the lower one:
\begin{equation}
    \pert{\tensor{g}{^\mu^\nu}}=-\tensor{\overline{g}}{^\mu^\alpha}\tensor{\overline{g}}{^\nu^\beta}\pert{\tensor{g}{_\alpha_\beta}}=-\tensor{\pertnb g}{^\mu^\nu},
\end{equation}
where in the second equality we introduced the tensor $\pertnb\tb{g}$ with components $\tensor{\pertnb g}{_\mu_\nu}\equiv\pert{\tensor{g}{_\mu_\nu}}$, and raise/lower its indices by means of the background. Note that $\tensor{\pertnb g}{_\mu_\nu}$ should be read as $\tensor{(\pertnb\tb{g})}{_\mu_\nu}$; the $\pertnb$ is now part of the tensor symbol, instead of an operator. Note that this introduction does \emph{not} imply $\pert{\tensor{g}{^\mu^\nu}}=\tensor{\pertnb g}{^\mu^\nu}$. The crux here is that moving indices does not commute with the perturbation operation $\pert{\,\dots}$. Namely, through the Leibniz rule this would incur additional perturbed factors of the metric in exactly the way as demonstrated above.

Thus, if we prescribe a first order perturbation $\tensor{\pertnb g}{_\mu_\nu}$, we have the two expansions
\begin{equation}
    \tensor{g}{_\mu_\nu}=\tensor{\overline{g}}{_\mu_\nu}+\epsilon\,\tensor{\pertnb g}{_\mu_\nu}+\order{\epsilon^2}\qand\tensor{g}{^\mu^\nu}=\tensor{\overline{g}}{^\mu^\nu}-\epsilon\,\tensor{\pertnb g}{^\mu^\nu}+\order{\epsilon^2}.
\end{equation}
From now on, we drop the overbar on the background metric, and assume that any appearance of the metric is the background one.

\subsection{The energy-momentum tensor}
Let us now turn to the r.h.s.\  of the Einstein equation \eqref{eq:einstein_eq_pert}: the energy-momentum tensor (EMT) of the system. According with \cite[\S5.1.2]{ellis_relativistic_2012}, we may write its decomposition in general form as
% \begin{equation}
%     \tensor*{T}{_\mu_\nu}=\rho\,\tensor{u}{_\mu}\tensor{u}{_\nu}+p\,\tensor{h}{_\mu_\nu}+2\,\tensor{q}{_{(\mu}}\tensor{u}{_{\nu)}}+\tensor*{\pi}{_\mu_\nu},\qq{where}\tensor{q}{^\mu}=\tensor{q}{^{\langle\mu\rangle}}\qand\tensor{\pi}{_\mu_\nu}=\tensor{\pi}{_{\langle\mu\nu\rangle}}
% \end{equation}
\begin{equation}\label{eq:EMT_def}
    \begin{split}
        \tensor{T}{_\mu_\nu}=(\tensor*{h}{^\alpha_\mu}-\tensor{u}{^\alpha}\tensor{u}{_\mu})(\tensor*{h}{^\beta_\nu}-\tensor{u}{^\beta}\tensor{u}{_\nu})\tensor{T}{_\alpha_\beta}&=\tensor{u}{^\alpha}\tensor{u}{^\beta}\tensor{T}{_\alpha_\beta}\,\tensor{u}{_\mu}\tensor{u}{_\nu}+\tfrac{1}{3}\tensor{h}{^\alpha^\beta}\tensor{T}{_\alpha_\beta}\,\tensor{h}{_\mu_\nu}-2\,\tensor{u}{^\alpha}\tensor{T}{_\alpha_\beta}\tensor*{h}{^\beta_{(\mu}}\tensor{u}{_{\nu)}}+\tensor{T}{_{\langle\mu\nu\rangle}} \\
        &=:\rho\,\tensor{u}{_\mu}\tensor{u}{_\nu}+p\,\tensor{h}{_\mu_\nu}+2\,\tensor{q}{_{(\mu}}\tensor{u}{_{\nu)}}+\tensor{\pi}{_\mu_\nu},
    \end{split}
\end{equation}
featuring the definitions
\begin{equation}
\tensor{u}{^\alpha}\tensor{u}{^\beta}\tensor{T}{_\alpha_\beta}=\rho,\quad\tensor{h}{^\alpha^\beta}\tensor*{T}{_\alpha_\beta}=3p,\quad\tensor{h}{^\mu^\alpha}\tensor{u}{^\beta}\tensor{T}{_\alpha_\beta}=-\tensor{q}{^\mu},\qand\tensor{T}{_{\langle\mu\nu\rangle}}=\tensor{\pi}{_\mu_\nu}.
\end{equation}
In the above, $\rho$ and $p$ are the familiar energy density and (relativistic) pressure, respectively, whereas $\tb{q}$ is the momentum density vector, and $\tb{\pi}$ is the anisotropic stress tensor. These latter two are in general necessary not exclusively at the perturbative level in order to encompass an anisotropic cosmology. We also note that, by construction, it holds that $\tensor{q}{^\mu}=\tensor{q}{^{\langle\mu\rangle}}$ and $\tensor{\pi}{_\mu_\nu}=\tensor{\pi}{_{\langle\mu\nu\rangle}}$. (For quantities with a single index, the angled brackets indicate simply projection onto the known hypersurface. That is, $\tensor{\mathfrak{v}}{^{\langle\mu\rangle}}\equiv\tensor*{h}{^\mu_\alpha}\tensor{\mathfrak{v}}{^\alpha}$.)

%We also emphasize that once $\rho$ and $p$ are defined by means of utilizing an arbitrary frame, they are invariant under diffeomorphisms of said frame. However, since a non-coordinate frame is not related to coordinate frame via a change of coordinates, the expressions \emph{will} change. In particular, expression independent of space in a non-coordinate frame formulation may become dependent on space in a coordinate one. As an example, suppose that we have two frames, each having chosen $\tb{u}$ as the timelike component, but with different frame choices for space: frame $\mathbb{C}$ is coordinated, and frame $\mathbb{N}$ is not. Then if the metric in frame $\mathbb{N}$ has metric components dependent only on time, expressing the metric in the coordinates of $\mathbb{C}$ will (generally) incur spatial dependence in the metric components. Supposing now that in $\mathbb{N}$ pressure $p$ depends only on time, this will mean that in $\mathbb{C}$ it \emph{obtains spatial dependence.} The reason is that $p$ is a density, and thus it will vary if the local volume element changes: $p$ having been space-independent in $\mathbb{N}$ can be interpreted by $\mathbb{C}$ as it still varying from place to place, but by the exact amount as $\mathbb{N}$'s volume changing (from $\mathbb{C}$'s perspective!) so as to be constant.

We also emphasize that once $\rho$ and $p$ are defined by means of an arbitrary frame, say $\mathbb{F}$, they are invariant under diffeomorphisms of $\mathbb{F}$. However, supposing that $\mathbb{F}$ is non-coordinate, it will not be related to a coordinate frame, say $\mathbb{G}$, via diffeomorphism. Therefore, the expressions may change, and in particular, they may pick up dependencies that were not there to begin with---concretely, if $\rho$ and $p$ defined in $\mathbb{F}$ depend only on cosmic time, they may also depend on space in $\mathbb{G}$. The reason is that these are \emph{densities,} and the volumes against which they are defined are the ones stemming from the metric. Since the metric, when expressed in $\mathbb{G}$, will pick up dependence on space, $\rho$ and $p$ will too. Physically, the reasoning is that the ``amount'' of energy/momentum located near an event is equal for both frames, but the local volume is not. Thus, their densities, as found through the EMT, will not be either.

The continuity equation for this general, decomposed EMT \eqref{eq:EMT_def} is
\begin{equation}
    0=\tensor{T}{_\mu_\alpha^{;\alpha}}=(\dot{\rho}+\theta(\rho+p)+\tensor{q}{^\alpha_{;\alpha}})\,\tensor{u}{_\mu}+\tensor{p}{_{|\mu}}+(\rho+p)\,\tensor{\dot{u}}{_\mu}+\tensor{\dot{q}}{_\mu}+\theta\,\tensor{q}{_\mu}+\tensor{q}{^\alpha}\tensor{u}{_{\mu|\alpha}}+\tensor{\pi}{_\mu_\alpha^{;\alpha}}.
\end{equation}
which we can also project along and perpendicular to the fluid flow to find
\begin{subequations}
    \begin{equation}
        0=-\tensor{u}{^\alpha}\tensor{T}{_\alpha_\beta^{;\beta}}=\dot{\rho}+\theta(\rho+p)+\tensor{q}{^\alpha_{;\alpha}}+\tensor{q}{_\alpha}\tensor{\dot{u}}{^\alpha}+\tensor{\sigma}{^\alpha^\beta}\tensor{\pi}{_\alpha_\beta}
    \end{equation}
    and
    \begin{equation}
        \tb{0}=\tensor{h}{^\mu^\alpha}\tensor{T}{_\alpha_\beta^{;\beta}}=\tensor{p}{^{|\mu}}+(\rho+p)\tensor{\dot{u}}{^\mu}+\tensor{\dot{q}}{^{\langle\mu\rangle}}+\theta\,\tensor{q}{^\mu}+\tensor{q}{_\alpha}\tensor{u}{^\mu^{|\alpha}}+\dot{u}{_\alpha}\tensor{\pi}{^\mu^\alpha}+\tensor{\pi}{^\mu^\alpha_{|\alpha}}.
    \end{equation}
\end{subequations}

We can then vary the EMT: similar to the metric, we introduce the expansion $\tensor{T}{_\mu_\nu}\to\tensor{T}{_\mu_\nu}+\epsilon\,\pert{\tensor{T}{_\mu_\nu}}+\order{\epsilon^2}$. By passing the perturbation to the defining quantities $\rho$, $p$, $\tb{q}$, and $\tb{\pi}$ of $\tb{T}$ \eqref{eq:EMT_def} (and substituting $\tb{h}$ in terms of $\tb{u}$ and $\tb{g}$ \eqref{eq:projection_op}), we find the following perturbation equation:
\begin{equation}
    \begin{split}
        \pert{\tensor{T}{_\mu_\nu}}&=(\pertnb\rho+\pertnb p)\,\tensor{u}{_\mu}\tensor{u}{_\nu}+2(\rho+p)\,\tensor{u}{_{(\mu}}\pert{\tensor{u}{_{\nu)}}}+p\,\tensor{\pertnb g}{_\mu_\nu}+\pertnb p\,\tensor{g}{_\mu_\nu}+2\,\pert{\tensor{q}{_{(\mu}}}\tensor{u}{_{\nu)}} \\
        &\qquad+2\,\tensor{q}{_{(\mu}}\pert{\tensor{u}{_{\nu)}}}+\pert{\tensor{\pi}{_\mu_\nu}}.
    \end{split}
\end{equation}
As in the previous subsection, we read this equation as an equality between \emph{perturbations of components} of tensors. We should like to work with \emph{components of perturbations,} however, for the perturbations are tensors and thus easier to handle mathematically. In order to go from the former to the latter, we make the definitions
\begin{equation}\label{eq:emt_pert_1}
    \tensor{\pertnb T}{_\mu_\nu}\equiv\pert{\tensor{T}{_\mu_\nu}},\quad\tensor{\pertnb u}{^\mu}\equiv\pert{\tensor{u}{^\mu}},\quad\tensor{\pertnb q}{^\mu}\equiv\pert{\tensor{q}{^\mu}},\qand\tensor{\pertnb\pi}{_\mu_\nu}\equiv\pert{\tensor{\pi}{_\mu_\nu}}.
\end{equation}
Just as with the metric tensor above, these definitions linking perturbations of components to components of perturbations represent the original objects and their perturbations. For instance, the perturbation of the vector field $\tb{u}$, $\pertnb\tb{u}$, is once more a vector field, and so its natural index position is up. From another point of view: when speaking of the vector field $\tb{u}$, $\tensor{u}{_\mu}$ is understood as equating $\tensor{u}{_\mu}\equiv\tensor{g}{_\mu_\alpha}\tensor{u}{^\alpha}$. Therefore, by Leibniz,
\begin{equation}
    \pert{\tensor{u}{_\mu}}=\pert{\tensor{g}{_\mu_\alpha}\tensor{u}{^\alpha}}=\tensor{g}{_\mu_\alpha}\pert{\tensor{u}{^\alpha}}+\pert{\tensor{g}{_\mu_\alpha}}\tensor{u}{^\alpha}=\tensor{\pertnb u}{_\mu}+\pert{\tensor{g}{_\mu_\alpha}}\tensor{u}{^\alpha}.
\end{equation}
Using this equation and similar ones for the other quantities in play, we are able to rewrite \eqref{eq:emt_pert_1} into
\begin{equation}
	\begin{split}
		\tensor{\pertnb T}{_\mu_\nu}&=(\pertnb\rho+\pertnb p)\tensor{u}{_\mu}\tensor{u}{_\nu}+2(\rho+p)\,\tensor{u}{_{(\mu}}\tensor{\pertnb u}{_{\nu)}}+2(\rho+p)\tensor{u}{_{(\mu}}\tensor{u}{^\alpha}\tensor{\pertnb g}{_{\nu)}_\alpha}+\pertnb p\,\tensor{g}{_\mu_\nu}+p\,\tensor{\pertnb g}{_\mu_\nu} \\
		&\qquad+2\,\tensor{\pertnb q}{_{(\mu}}\tensor{u}{_{\nu)}}+2\,\tensor{q}{_{(\mu}}\tensor{\pertnb u}{_{\nu)}}+2\,\tensor{q}{^\alpha}\tensor{u}{_{(\mu}}\tensor{\pertnb g}{_{\nu)}_\alpha}+2\,\tensor{u}{^\alpha}\tensor{q}{_{(\mu}}\tensor{\pertnb g}{_{\nu)}_\alpha}+\tensor{\pertnb\pi}{_\mu_\nu}.
	\end{split}
\end{equation}
We now wish to project this equation along and perpendicular to the fluid flow, as that will allow us to extract the perturbed components of interest. To do so efficiently, we also employ some identities between the perturbations that follow from perturbing known relations. For instance,
\begin{equation}
    \tensor{u}{_\alpha}\tensor{u}{^\alpha}=-1\implies0=\tensor{u}{_\alpha}\pert{\tensor{u}{^\alpha}}+\pert{\tensor{u}{_\alpha}}\tensor{u}{^\alpha}=2\,\tensor{u}{_\alpha}\tensor{\pertnb u}{^\alpha}+\tensor{u}{^\alpha}\tensor{u}{^\beta}\tensor{\pertnb g}{_\alpha_\beta}.
\end{equation}
Similarly we use the relations
\begin{equation}
    \tensor{u}{_\alpha}\tensor{\pertnb q}{^\alpha}+\tensor{q}{_\alpha}\tensor{\pertnb u}{^\alpha}+\tensor{q}{^\alpha}\tensor{u}{^\beta}\tensor{\pertnb g}{_\alpha_\beta}=0,\quad\tensor{u}{^\alpha}\tensor{\pertnb\pi}{_\mu_\alpha}=-\tensor{\pi}{_\mu_\alpha}\tensor{\pertnb u}{^\alpha}.
\end{equation}
We can then find the projections as follows:
\begin{subequations}\label{eq:EMTExpansions}
    \begin{align}
        \tensor{u}{^\alpha}\tensor{u}{^\beta}\tensor{\pertnb T}{_\alpha_\beta}&=\pertnb\rho-\rho\,\tensor{u}{^\alpha}\tensor{u}{^\beta}\tensor{\pertnb g}{_\alpha_\beta}+2\,\tensor{q}{_\alpha}\tensor{\pertnb u}{^\alpha}, \\
        \tensor{h}{^\alpha^\beta}\tensor{\pertnb T}{_\alpha_\beta}&=3\,\pertnb p+2\,\tensor{q}{_\alpha}\tensor{\pertnb u}{^\alpha}+(\tensor{T}{^\alpha^\beta}-\rho\,\tensor{u}{^\alpha}\tensor{u}{^\beta})\tensor{\pertnb g}{_\alpha_\beta}, \\
        \tensor{u}{^\alpha}\tensor*{h}{^\mu^\beta}\tensor{\pertnb T}{_\alpha_\beta}&=-\tensor{\pertnb q}{^{\langle\mu\rangle}}-[(\rho+p)\,\tensor*{h}{^\mu_\alpha}+\tensor{u}{_\alpha}\tensor{q}{^\mu}+\tensor*{\pi}{^\mu_\alpha}]\tensor{\pertnb u}{^\alpha}-\tensor{h}{^\mu^\alpha}(\rho\,\tensor{u}{^\beta}+\tensor{q}{^\beta})\tensor{\pertnb g}{_\alpha_\beta}, \\
        \tensor{\pertnb T}{_{\langle\mu\nu\rangle}}&=\tensor{\pertnb\pi}{_{\langle\mu\nu\rangle}}+p\,\tensor{\pertnb g}{_{\langle\mu\nu\rangle}}+2\,\tensor{q}{_{\langle\mu}}(\tensor{\pertnb u}{_{\nu\rangle}}+\tensor{u}{^\alpha}\tensor{\pertnb g}{_{\nu\rangle}_\alpha}).
    \end{align}
\end{subequations}

We note that from these equations, we can isolate the first order perturbations $\pertnb\rho$, $\pertnb p$, $\tensor{\pertnb q}{^{\langle\mu\rangle}}$, and $\tensor{\pertnb\pi}{_{\langle\mu\nu\rangle}}$ and hence find expressions for them involving the background parameters, and the metric and velocity perturbations. As a sanity check, we find that upon inputting conformal flat FLRW, $\tensor{g}{_\mu_\nu}=a^2\tensor{\eta}{_\mu_\nu}$, with fluid flow $\tensor{u}{^\mu}=a^{-1}\tensor*{\delta}{^\mu_0}$, the equations reduce to the standard interpretation of perturbed EMT components cf.\ \cite[\S7.3]{mukhanov_physical_2005}.

\subsection{The Einstein tensor \& Einstein equation}
The variation in the Einstein tensor is trickier to find, since it involves covariant differentiation of quantities. By virtue of the metric being perturbed, the Christoffel symbols are also perturbed and hence covariant differentiation becomes more involved. Although an analytical approach is possible via the Palatini identity for the perturbations of the Ricci tensor \cite[\S35.13]{misner_gravitation_2017}, we instead utilize the Mathematica package \texttt{xPand} \cite{pitrou_xpand_2013}, based on \texttt{xAct}, in order to produce a ``raw'' first order perturbation to the Einstein tensor. We then massage the resulting expression, eventually ending up with the following perturbation equation:
% \begin{equation}
%     \begin{split}
%         \tensor{\delta G}{_\mu_\nu}&=\tfrac{1}{2}\tensor{\delta g}{_\alpha_\beta}\left(\tensor{g}{_\mu_\nu}\tensor{R}{^\alpha^\beta}-\tensor*{\delta}{^\alpha_\mu}\tensor*{\delta}{^\beta_\nu}R\right)+\tensor*{\delta g}{_\alpha_\beta_{;\gamma\delta}}\left(\tensor{g}{^\alpha^\delta}\tensor*{\delta}{^\beta_{(\mu}}\tensor*{\delta}{^\gamma_{\nu)}}-\tfrac{1}{2}\tensor{g}{^\alpha^\delta}\tensor{g}{^\beta^\gamma}\tensor{g}{_\mu_\nu}\right) \\
%         &\qquad+\tfrac{1}{2}\tensor{g}{_\mu_\nu}\square(\Tr\delta g)-\tfrac{1}{2}\square\tensor{\delta g}{_\mu_\nu}-\tfrac{1}{2}\tensor{(\Tr\delta g)}{_{;\mu\nu}}.
%     \end{split}
% \end{equation}
% \begin{equation}
%     \begin{split}
%         \tensor{\delta G}{^\mu_\nu}&=\tfrac{1}{2}\left[\square(\Tr\delta g)+\tensor{R}{^\alpha^\beta}\tensor{\delta g}{_\alpha_\beta}-\tensor{g}{^\alpha^\delta}\tensor{g}{^\beta^\gamma}\tensor{\delta g}{_\alpha_\beta_{;\gamma\delta}}\right]\tensor*{\delta}{^\mu_\nu}+\tensor{g}{^\alpha^\delta}\tensor{g}{^\mu^{(\beta}}\tensor*{\delta}{^{\gamma)}_\nu}\tensor{\delta g}{_\alpha_\beta_{;\gamma\delta}} \\
%         &\quad-\tfrac{1}{2}R\tensor{g}{^\alpha^\mu}\tensor*{\delta}{^\beta_\nu}\tensor{\delta g}{_\alpha_\beta}-\tfrac{1}{2}\tensor{g}{^\mu^\alpha}\,\square(\tensor{\delta g}{_\alpha_\nu})-\tfrac{1}{2}\tensor*{(\Tr\delta g)}{^{;\mu}_{;\nu}}.
%     \end{split}
% \end{equation}
\begin{equation}
    \begin{split}
        \pert{\tensor{G}{_\mu_\nu}}\equiv\tensor{\pertnb G}{_\mu_\nu}&=\tfrac{1}{2}\left[\tensor{R}{^\alpha^\beta}\tensor{\pertnb g}{_\alpha_\beta}-\tensor{\pertnb g}{_\alpha_\beta^{;\alpha\beta}}+\square(\Tr\pertnb g)\right]\tensor{g}{_\mu_\nu}-\tfrac{1}{2}R\,\tensor{\pertnb g}{_\mu_\nu}-\tfrac{1}{2}\square\tensor{\pertnb g}{_\mu_\nu}\\
        &\qquad-\tfrac{1}{2}\tensor{(\Tr\pertnb g)}{_{;\mu\nu}}+\tensor*{\pertnb g}{^\alpha_{(\mu}_{;\nu)\alpha}},
    \end{split}
\end{equation}
where we defined $\Tr\pertnb g=\tensor*{\pertnb g}{^\alpha_\alpha}$. Its projections are given by
\begin{subequations}\label{eq:EinsteinExpansions}
    \begin{align}
		\begin{split}
			\tensor{u}{^\alpha}\tensor{u}{^\beta}\tensor{\pertnb G}{_\alpha_\beta}&=\tensor{u}{^\alpha}\tensor{u}{^\beta}(\tensor*{\pertnb g}{^\gamma_\alpha_{;\beta\gamma}}-\tfrac{1}{2}\square\tensor{\pertnb g}{_\alpha_\beta}-\tfrac{1}{2}R\,\tensor{\pertnb g}{_\alpha_\beta})+\tfrac{1}{2}\tensor{\pertnb g}{_\alpha_\beta^{;\alpha\beta}}-\tfrac{1}{2}\tensor{R}{^\alpha^\beta}\tensor{\pertnb g}{_\alpha_\beta}-\tfrac{1}{2}\square(\Tr\pertnb g)\\
			&\qquad-\tfrac{1}{2}\ddot{(\Tr\pertnb g)}+\tfrac{1}{2}\tensor{\dot{u}}{^\alpha}\tensor{(\Tr\pertnb g)}{_{;\alpha}},
		\end{split} \\
		\begin{split}
			\tensor{h}{^\alpha^\beta}\tensor{\pertnb G}{_\alpha_\beta}&=\tensor{u}{^\alpha}\tensor{u}{^\beta}(\tensor*{\pertnb g}{^\gamma_\alpha_{;\beta\gamma}}-\tfrac{1}{2}\square\tensor{\pertnb g}{_\alpha_\beta}-\tfrac{1}{2}R\,\tensor{\pertnb g}{_\alpha_\beta})-\tfrac{1}{2}\tensor{\pertnb g}{_\alpha_\beta^{;\alpha\beta}}+\tfrac{3}{2}\tensor{R}{^\alpha^\beta}\tensor{\pertnb g}{_\alpha_\beta}+\tfrac{1}{2}\square(\Tr\pertnb g)\\
			&\qquad-\tfrac{1}{2}\ddot{(\Tr\pertnb g)}+\tfrac{1}{2}\tensor{\dot{u}}{^\alpha}\tensor{(\Tr\pertnb g)}{_{;\alpha}}-\tfrac{1}{2}R\,\Tr\pertnb g,
		\end{split} \\
		\begin{split}
			\tensor{u}{^\alpha}\tensor*{h}{^\beta_\mu}\tensor{\pertnb G}{_\alpha_\beta}&=\tensor{u}{^\alpha}\tensor*{h}{^\beta_\mu}(\tensor*{\pertnb g}{^\gamma_{(\alpha;\beta)}_\gamma}-\tfrac{1}{2}\square\tensor{\pertnb g}{_\alpha_\beta}-\tfrac{1}{2}R\,\tensor{\pertnb g}{_\alpha_\beta})-\tfrac{1}{2}\tensor{\dot{(\Tr\pertnb g)}}{_{|\mu}}+\tfrac{1}{2}\tensor{u}{^\alpha_{|\mu}}\tensor{(\Tr\pertnb g)}{_{|\alpha}},
		\end{split} \\
		\begin{split}
			\tensor{\pertnb G}{_{\langle\mu\nu\rangle}}&=(\tensor*{h}{^\alpha_\mu}\tensor*{h}{^\beta_\nu}-\tfrac{1}{3}\tensor{h}{^\alpha^\beta}\tensor{h}{_\mu_\nu})(\tensor*{\pertnb g}{^\gamma_{(\alpha;\beta)}_\gamma}-\tfrac{1}{2}\square\tensor{\pertnb g}{_\alpha_\beta})-\tfrac{1}{2}R\,\tensor{\pertnb g}{_{\langle\mu\nu\rangle}}-\tfrac{1}{2}\tensor{(\Tr\pertnb g)}{_{|(\mu\nu)}} \\
			&\qquad+\tfrac{1}{2}\dot{(\Tr\pertnb g)}\,\tensor{\sigma}{_\mu_\nu}+\tfrac{1}{6}\triangle(\Tr\pertnb g)\,\tensor{h}{_\mu_\nu}.
		\end{split}
	\end{align}
\end{subequations}
With the expansions of the EMT \eqref{eq:EMTExpansions} and the Einstein tensor \eqref{eq:EinsteinExpansions}, we are now in a position to formulate the perturbed Einstein equations per component. These are
\begin{equation}
    \tensor{\pertnb G}{_\mu_\nu}+\Lambda\,\tensor{\pertnb g}{_\mu_\nu}=\kappa\,\tensor{\pertnb T}{_\mu_\nu}\iff\begin{cases}\medskip\tensor{u}{^\alpha}\tensor{u}{^\beta}\tensor{\pertnb G}{_\alpha_\beta}+\Lambda\,\tensor{u}{^\alpha}\tensor{u}{^\beta}\tensor{\pertnb g}{_\alpha_\beta}=\kappa\,\tensor{u}{^\alpha}\tensor{u}{^\beta}\tensor{\pertnb T}{_\alpha_\beta} \\ \medskip \tensor{h}{^\alpha^\beta}\tensor{\pertnb G}{_\alpha_\beta}+\Lambda\,\tensor{h}{^\alpha^\beta}\tensor{\pertnb g}{_\alpha_\beta}=\kappa\,\tensor{h}{^\alpha^\beta}\tensor{\pertnb T}{_\alpha_\beta} \\ \medskip\tensor{h}{^\mu^\alpha}\tensor{u}{^\beta}\tensor{\pertnb G}{_\alpha_\beta}+\Lambda\,\tensor{h}{^\mu^\alpha}\tensor{u}{^\beta}\tensor{\pertnb g}{_\alpha_\beta}=\kappa\,\tensor{h}{^\mu^\alpha}\tensor{u}{^\beta}\tensor{\pertnb T}{_\alpha_\beta} \\ \tensor{\pertnb G}{_{\langle\mu\nu\rangle}}+\Lambda\,\tensor{\pertnb g}{_{\langle\mu\nu\rangle}}=\kappa\,\tensor{\pertnb T}{_{\langle\mu\nu\rangle}}\end{cases},
\end{equation}
wherein now all the terms are known. Explicitly, we have that the perturbation equations of $\rho$, $p$, $\tb{q}$, and $\tb{\pi}$, are given by
\begin{subequations}\label{eq:efes_pert_eqs}
	\begin{align}
		\begin{split}
			\kappa\,\pertnb\rho&=\tensor{u}{^\alpha}\tensor{u}{^\beta}(\tensor*{\pertnb g}{^\gamma_\alpha_{;\beta\gamma}}-\tfrac{1}{2}\square\tensor{\pertnb g}{_\alpha_\beta}-\tfrac{1}{2}(R-2\Lambda-2\kappa\rho)\,\tensor{\pertnb g}{_\alpha_\beta})-2\kappa\,\tensor{q}{_\alpha}\tensor{\pertnb u}{^\alpha}+\tfrac{1}{2}\tensor{\pertnb g}{_\alpha_\beta^{;\alpha\beta}} \\
			&\qquad-\tfrac{1}{2}\ddot{(\Tr\pertnb g)} -\tfrac{1}{2}\tensor{R}{^\alpha^\beta}\tensor{\pertnb g}{_\alpha_\beta}-\tfrac{1}{2}\square(\Tr\pertnb g)+\tfrac{1}{2}\tensor{\dot{u}}{^\alpha}\tensor{(\Tr\pertnb g)}{_{;\alpha}},
		\end{split} \\
		\begin{split}
			3\kappa\,\pertnb p&=\tensor{u}{^\alpha}\tensor{u}{^\beta}(\tensor*{\pertnb g}{^\gamma_\alpha_{;\beta\gamma}}-\tfrac{1}{2}\square\tensor{\pertnb g}{_\alpha_\beta}-\tfrac{1}{2}(R-2\Lambda-2\kappa\rho)\,\tensor{\pertnb g}{_\alpha_\beta})-2\kappa\,\tensor{q}{_\alpha}\tensor{\pertnb u}{^\alpha}-\tfrac{1}{2}\tensor{\pertnb g}{_\alpha_\beta^{;\alpha\beta}} \\
			&\qquad-\tfrac{1}{2}\ddot{(\Tr\pertnb g)}+\tfrac{1}{2}\tensor{R}{^\alpha^\beta}\tensor{\pertnb g}{_\alpha_\beta}+\tfrac{1}{2}\square(\Tr\pertnb g)+\tfrac{1}{2}\tensor{\dot{u}}{^\alpha}\tensor{(\Tr\pertnb g)}{_{;\alpha}},
		\end{split} \\
		\begin{split}
			\kappa\,\tensor{\pertnb q}{^{\langle\mu\rangle}}&=\tfrac{1}{2}\tensor{\dot{(\Tr\pertnb g)}}{^{|\mu}}-\tensor{u}{^\alpha}\tensor*{h}{^\beta^\mu}(\tensor*{\pertnb g}{^\gamma_{(\alpha}_{;\beta)}_\gamma}-\frac{1}{2}\square\tensor{\pertnb g}{_\alpha_\beta}-\tfrac{1}{2}R\,\tensor{\pertnb g}{_\alpha_\beta})-\tfrac{1}{2}\tensor{u}{^\alpha^{|\mu}}\tensor{(\Tr\pertnb g)}{_{|\alpha}} \\
			&\qquad-(\rho\,\tensor{u}{^\alpha}+\tensor{q}{^\alpha})\tensor*{h}{^\beta^\mu}\tensor{\pertnb g}{_\alpha_\beta}-((\rho+p)\tensor*{h}{^\mu_\alpha}+\tensor{u}{_\alpha}\tensor{q}{^\mu}+\tensor*{\pi}{^\mu_\alpha})\tensor{\pertnb u}{^\alpha},
		\end{split} \\
		\begin{split}
			\kappa\,\tensor{\pertnb\pi}{_{\langle\mu\nu\rangle}}&=(\tensor*{h}{^\alpha_\mu}\tensor*{h}{^\beta_\nu}-\tfrac{1}{3}\tensor{h}{^\alpha^\beta}\tensor{h}{_\mu_\nu})(\tensor*{\pertnb g}{^\gamma_{(\alpha;\beta)}_\gamma}-\tfrac{1}{2}\square\tensor{\pertnb g}{_\alpha_\beta})-\tfrac{1}{2}(R-2\Lambda+2\kappa p)\,\tensor{\pertnb g}{_{\langle\mu\nu\rangle}} \\
			&\qquad-\tfrac{1}{2}\tensor{(\Tr\pertnb g)}{_{|(\mu\nu)}}+\tfrac{1}{2}\dot{(\Tr\pertnb g)}\,\tensor{\sigma}{_\mu_\nu}+\tfrac{1}{6}\triangle(\Tr\pertnb g)\,\tensor{h}{_\mu_\nu}-2\kappa\,\tensor{q}{_{\langle\mu}}(\tensor{\pertnb u}{_{\nu\rangle}}+\tensor{u}{^\alpha}\tensor{\pertnb g}{_{\nu\rangle}_\alpha}).
		\end{split}
	\end{align}
\end{subequations}

\section{Fixing the Newtonian gauge}\label{sec:newton_gauge}
%Up to now our treatment has been entirely general: that is to say, it holds for any metric, metric frame, and choice of perturbation. (We only fixed the conditions on the fluid flow 4-vector $\tensor{u}{^\mu}$.) We shall now choose to evaluate a scalar perturbation in the Newtonian gauge, and find the perturbed Einstein equation that belongs to it. We do this as this seems closest to our desired application to structure formation in cosmology, and what is done commonly in the literature. As we noted above, since our theory is not gauge invariant (we do not explicitly know the gauge transformation laws, and hence cannot formulate a gauge-invariant theory), our choice of Newtonian gauge is specific. 

Up to now, our treatment has been entirely general. That is to say, it holds for any metric, metric frame, and choice of perturbation. (We only fixed the conditions on the fluid flow 4-vector $\boldsymbol{u}$, see \S\ref{sec:3+1}.) We shall now evaluate these equations for perturbations \emph{fixed to the Newtonian gauge.} We do this for two reasons: firstly, this is closest to our desired application to CMB analysis and structure formation; and secondly, the metric theory of gauge transformations and invariance is complicated to such an extent that it would cloud any useful conclusions to be drawn from this approach. Much research on cosmological gauge theory, and with an eye towards application in homogeneous, anisotropic spacetimes, has been done, and can be found in e.g. \cite{nakamura_construction_2013, tsagas_relativistic_2008}.

% In the first subsection, we give a small discussion on the complexity of gauge transformations in our more general setting, though we will leave a complete treatment to future works. Thereafter, we proceed with evaluating scalar perturbations, and derive (under some simplifying assumptions) the equivalent of the Mukhanov-Sasaki equation for our setting. Finally, we briefly evaluate the pure tensor mode perturbations in this universe, as this may prove useful for gravitational wave analysis.
We first proceed with evaluating pure scalar perturbations in the Newtonian gauge. A thermodynamical argument then allows us to link the perturbations in the density and pressure, in order to give one equation that must be satisfied by the perturbation at all events. This will be the equivalent of the Mukhanov-Sasaki equation for our setting of homogeneous, anisotropic spacetimes. We then proceed with evaluating pure tensor mode perturbations, as such an analysis and understanding may prove useful for gravitational wave analyses in anisotropic spacetimes.

\subsection{Scalar perturbations}\label{sec:scalarPerts}
We now fix the Newtonian gauge, considering only scalar perturbations. Thus, we assume that the perturbation to the metric, $\tensor{\pertnb g}{_\mu_\nu}$, is given by
\begin{equation}\label{eq:pert-newton-gauge}
    \tensor{\pertnb g}{_\mu_\nu}\equiv-2\phi\,\tensor{u}{_\mu}\tensor{u}{_\nu}-2\psi\,\tensor{h}{_\mu_\nu}=-2(\phi+\psi)\,\tensor{u}{_\mu}\tensor{u}{_\nu}-2\psi\,\tensor{g}{_\mu_\nu};
\end{equation}
in other words, we set $B=E=\tb{S}=\tb{F}=\tb{T}=0$. We then want to fill in this perturbation into the perturbation equations of the Einstein tensor in equation \eqref{eq:EinsteinExpansions}. Since this is quite a laborious process, we once more utilize \texttt{xPand} in order to fill in this for $\tensor{\pertnb g}{_\mu_\nu}$ and expand, and then we simplify by hand using known relations such as the Raychaudhuri equation. The result is the following set of equations:
\begin{subequations}\label{eq:perturbedEinsteinComponents}
    \begin{align}
        \tensor{u}{^\alpha}\tensor{u}{^\beta}\tensor{\pertnb G}{_\alpha_\beta}&=2\lozenge\psi+2(\phi+\psi)(\tensor{u}{^\alpha}\tensor{u}{^\beta}\tensor{G}{_\alpha_\beta}-\tfrac{1}{3}\theta^2+\sigma^2+3\omega^2) \\
        \tensor{h}{^\alpha^\beta}\tensor{\pertnb G}{_\alpha_\beta}&=6\ddot{\psi}+2\lozenge(\phi-\psi)+2\tensor{\dot{u}}{^\alpha}\tensor{(2\phi-\psi)}{_{;\alpha}}+4\theta(\dot{\phi}+\dot{\psi})+(\phi+\psi)(4\dot{\theta}+2\theta^2+6\sigma^2+2\omega^2) \\
        \tensor{h}{^\mu^\alpha}\tensor{u}{^\beta}\tensor{\pertnb G}{_\alpha_\beta}&=2\tensor{\dot{\psi}}{^{|\mu}}+\tfrac{1}{3}\theta\tensor{(2\phi+\psi)}{^{|\mu}}+3\tensor{\omega}{^\alpha^\mu}\tensor{\phi}{_{;\alpha}}-\tensor{\sigma}{^\mu^\alpha}\tensor{(\phi+2\psi)}{_{;\alpha}}+2(\phi+\psi)(\tensor{\omega}{^\alpha^\mu_{|\alpha}}+2\tensor{\dot{u}}{_\alpha}\tensor{\omega}{^\alpha^\mu}) \\
        \begin{split}
            \tensor{\pertnb G}{_{\langle\mu\nu\rangle}}&=\tensor{(\psi-\phi)}{_{\langle;\mu\nu\rangle}}+\tensor{\dot{u}}{_{\langle\mu}}\tensor{(\phi+\psi)}{_{;\nu\rangle}}-2(\dot{\phi}+\dot{\psi})\tensor{\sigma}{_\mu_\nu} \\
            &\qquad+2(\phi+\psi)\left(\tensor{\omega}{_{\langle\mu}^\alpha}\tensor{u}{_{\nu\rangle}_{;\alpha}}+2\tensor{\dot{u}}{^\alpha}\tensor{\sigma}{_\alpha_{(\mu}}\tensor{u}{_{\nu)}}-\tensor{\dot{\sigma}}{_\mu_\nu}-\theta\,\tensor{\sigma}{_\mu_\nu}\right)
        \end{split}
    \end{align}
\end{subequations}

We can then write the perturbed Einstein equations for a scalar perturbation in Newtonian gauge. Combining the components of the perturbed Einstein tensor \eqref{eq:perturbedEinsteinComponents} with those of the EMT \eqref{eq:EMTExpansions} (populated with the perturbation form \eqref{eq:pert-newton-gauge}), we can derive the following perturbation equations:
\begin{myEquationFrame}[frametitle = \colorbox{white}{Perturbation equations for scalar perturbations in Newtonian gauge}]
    \begin{subequations}\label{eq:friedmann-perts-newton}
        \begin{align}
            \kappa\,\pertnb\rho&=2\lozenge\psi+2(\phi+\psi)\left(\tensor{G}{_\alpha_\beta}\tensor{u}{^\alpha}\tensor{u}{^\beta}-\tfrac{1}{3}\theta^2+\sigma^2+3\omega^2\right)-2\phi(\kappa\rho+\Lambda)-2\kappa\,\tensor{q}{_\alpha}\tensor{\pertnb u}{^\alpha} \label{eq:densityPerturbation}\\
            \begin{split}\label{eq:pressurePerturbation} 
                3\kappa\,\pertnb p&=6\ddot{\psi}+2\lozenge(\phi-\psi)+2\tensor{\dot{u}}{^\alpha}\tensor{(2\phi-\psi)}{_{;\alpha}}+4\theta(\dot{\phi}+\dot{\psi})+(\phi+\psi)\left(4\dot{\theta}+2\theta^2+6\sigma^2+2\omega^2\right) \\
                &\qquad+6\psi(\kappa p-\Lambda)-2\kappa\,\tensor{q}{_\alpha}\tensor{\pertnb u}{^\alpha}
            \end{split} \\
            \begin{split}
                \kappa\,\tensor{\pertnb q}{^{\langle\mu\rangle}}&=-2\tensor{\dot{\psi}}{^{|\mu}}-\tfrac{1}{3}\theta\tensor{(2\phi+\psi)}{^{|\mu}}+3\tensor{\omega}{^\mu^\alpha}\tensor{\phi}{_{;\alpha}}+\tensor{\sigma}{^\mu^\alpha}\tensor{(\phi+2\psi)}{_{;\alpha}}-2(\phi+\psi)(\tensor{\omega}{^\alpha^\omega_{|\alpha}}+2\tensor{\dot{u}}{_\alpha}\tensor{\omega}{^\alpha^\mu} \\
                &\qquad-\kappa(\rho+p)\,\tensor{\pertnb u}{^{\langle\mu\rangle}}-\kappa(\phi+p)\tensor{q}{^\mu}-\kappa\,\tensor*{\pi}{^\mu_\alpha}\tensor{\pertnb u}{^\alpha}
            \end{split} \\
            \begin{split}\label{eq:aniStressTensorPerturbation}
                \kappa\,\tensor{\pertnb\pi}{_{\langle\mu\nu\rangle}}&=-\kappa\,\tensor{q}{_{\langle\mu}}\tensor{\pertnb u}{_{\nu\rangle}}-\kappa\phi\,\tensor{q}{_{(\mu}}\tensor{u}{_{\nu)}}+\tensor{(\psi-\phi)}{_{\langle;\mu\nu\rangle}}+\tensor{\dot{u}}{_{\langle\mu}}\tensor{(\phi+\psi)}{_{;\nu\rangle}}-2(\dot{\phi}+\dot{\psi})\tensor{\sigma}{_\mu_\nu} \\
                &\qquad+2(\phi+\psi)\left(\tensor{\omega}{_{\langle\mu}^\alpha}\tensor{u}{_{\nu\rangle}_{;\alpha}}+2\tensor{\dot{u}}{^\alpha}\tensor{\sigma}{_\alpha_{(\mu}}\tensor{u}{_{\nu)}}-\tensor{\dot{\sigma}}{_\mu_\nu}-\theta\,\tensor{\sigma}{_\mu_\nu}\right)
            \end{split}
            % \begin{split}
            %     \kappa\,\tensor*{\delta\pi}{^\mu_\nu}&=\tensor{u}{^\mu}\tensor{\pi}{_\nu_\alpha}\tensor{\delta u}{^\alpha}-\left(\tensor{q}{^\mu}\tensor{h}{_\nu_\alpha}+\tensor{q}{_\nu}\tensor*{h}{^\mu_\alpha}-\tfrac{2}{3}\tensor{q}{_\alpha}\tensor*{h}{^\mu_\nu}\right)\tensor{\delta u}{^\alpha}-2(\phi+\psi)\tensor{R}{^\mu_\alpha_\beta_\nu}\tensor{u}{^\alpha}\tensor{u}{^\beta} \\
            %     &\qquad+\tfrac{1}{3}\tensor*{h}{^\mu_\nu}\left[\lozenge(\phi-\psi)+2\tensor{\dot{u}}{^\alpha}\tensor{(\phi+\psi)}{_{;\alpha}}+2\theta(\dot{\phi}+\dot{\psi})+(\phi+\psi)\left(2\dot{\theta}+2\theta^2-8\omega^2\right)\right] \\
            %     &\qquad-2\tensor{h}{^\mu^{(\alpha}}\tensor*{h}{^{\beta)}_\nu}\left[(\dot{\phi}+\dot{\psi})\tensor{u}{_\alpha_{;\beta}}+\tensor{\dot{u}}{_\alpha}\tensor{(\phi+\psi)}{_{;\beta}}+[\tensor{\dot{u}}{_\alpha_{;\beta}}+\theta\tensor{u}{_\alpha_{;\beta}}-\tfrac{1}{2}\tensor{u}{_\alpha_{;\gamma}}\tensor{u}{_\beta^{;\gamma}}](\phi+\psi)\right]
            % \end{split}
        \end{align}
    \end{subequations}
\end{myEquationFrame}

Finally, we wish to combine the equations above, particularly \eqref{eq:densityPerturbation} and \eqref{eq:pressurePerturbation}, in order to arrive at an equation that parallels the ``standard'' equation for scalar perturbation theory---found e.g.\ in \cite[\S10.2]{ellis_relativistic_2012}---as announced above. In order to keep as much generalization open as possible, we do not a priori assume an isentropic perturbation. The relation between pressure and density perturbations is thus given by \cite[\S7.5]{mukhanov_physical_2005}
\begin{equation}\label{eq:soundwaves}
    \pertnb p=c_s^2\,\pertnb\rho+\tau\delta S,\qq{where}c_s:=\sqrt{\left(\pdv{p}{\rho}\right)_S}\qand\tau:=\left(\pdv{p}{S}\right)_\rho,
\end{equation}
where $\delta S$ represents the change in entropy due to the perturbation. In the above, $c_s$ has the interpretation of the speed of sound in the medium, and $\tau$ of a temperature in the therodynamic sense. Populating this relation \eqref{eq:soundwaves} with \eqref{eq:densityPerturbation} and \eqref{eq:pressurePerturbation} and shuffling some terms around, we find the relation
% \begin{equation}
%     \begin{split}
%         \tfrac{1}{3}\kappa(1-3c_s^2)\tensor{q}{_\alpha}\tensor{\delta u}{^\alpha}+\kappa\tau\,\delta S&=\ddot{\psi}-c_s^2\lozenge\psi+\tfrac{1}{3}\lozenge(\phi-\psi)+\tfrac{1}{3}\tensor{\dot{u}}{^\alpha}\tensor{(2\phi-\psi)}{_{;\alpha}}+\tfrac{2}{3}\theta(\dot{\phi}+\dot{\psi}) \\
%         &\qquad+(\phi+\psi)\left[\tfrac{2}{3}\dot{\theta}+\tfrac{1}{3}(1+c_s^2)\theta^2+(1-c_s^2)\sigma^2+\tfrac{1}{3}(1-9c_s^2)\omega^2-c_s^2(\tfrac{1}{2}R+\tensor{R}{_\alpha_\beta}\tensor{u}{^\alpha}\tensor{u}{^\beta})\right]
%     \end{split}
% \end{equation}
% \begin{equation}
%     \begin{split}
%         &\ddot{\psi}-c_s^2\lozenge\psi+\tfrac{1}{3}\lozenge(\phi-\psi)+\tfrac{1}{3}\tensor{\dot{u}}{^\alpha}\tensor{(2\phi-\psi)}{_{;\alpha}}+\tfrac{2}{3}\theta(\dot{\phi}+\dot{\psi})+\Lambda(c_s^2\phi-\psi)+\kappa(\psi p+c_s^2\phi\rho)-\tfrac{1}{3}\kappa\,\tensor{q}{^\alpha}\pert{\tensor{u}{_\alpha}} \\ 
%         &\qquad+c_s^2\kappa\,\tensor{q}{_\alpha}\tensor{\pertnb u}{^\alpha}+(\phi+\psi)\left(\tfrac{2}{3}\dot{\theta}+\tfrac{1}{3}\theta^2(1+c_s^2)+(1-c_s^2)\sigma^2+\tfrac{1}{3}(1-9c_s^2)\omega^2-c_s^2\tensor{u}{^\alpha}\tensor{u}{^\beta}\tensor{G}{_\alpha_\beta}\right)+\tfrac{1}{2}\kappa\tau\,\delta S=0.
%     \end{split}
% \end{equation}
\begin{equation}
    \begin{split}
        3\kappa\tau\,\delta S&=6\ddot{\psi}-6c_s^2\lozenge\psi+2\lozenge(\phi-\psi)+2\tensor{\dot{u}}{^\alpha}\tensor{(2\phi-\psi)}{_{;\alpha}}+4\theta(\dot{\phi}+\dot{\psi})+6\psi\kappa(p-c_s^2\rho)-6\psi\Lambda(1+c_s^2) \\
        &\qquad+2(\phi+\psi)\left(2\dot{\theta}+(1+c_s^2)\theta^2+3(1-c_s^2)\sigma^2+(1-9c_s^2)\omega^2\right)+2\kappa(3c_s^2-1)\,\tensor{q}{_\alpha}\tensor{\pertnb u}{^\alpha}.
    \end{split}
\end{equation}
This is quite the equation to consider, so let us make some simplifying assumptions in order to increase its tractability.
\begin{enumerate}
    \item \emph{The perturbation is isentropic.} That is to say, $\delta S\equiv0$. Note that this assumption is slightly stronger than assuming that the perturbation is adiabatic---see \cite[\S2.2]{bertschinger_cosmological_1995} for comments on the distinction.
    \item \emph{No pure scalar contributions to stress.} Eyeing \eqref{eq:aniStressTensorPerturbation}, this means $\phi=\psi$. Alternatively, this assumption can be stated as the weak-field approximation to GR that should lead to Newtonian gravity to first order. Then $\psi$ (or $\phi$) represents the Newtonian potential \cite[\S4.4]{wald_general_1984}.
    \item \emph{The equation of state is given by $p=c_s^2\rho$.} If the sound speed $c_s^2$ is constant, like in epochs with component domination, then this is necessarily true. Otherwise the equation of state may vary, see \cite[\S3.2]{tsagas_relativistic_2008}.
\end{enumerate}
Applying these assumptions, we arrive at what we term the \emph{homogeneous \& anisotropic, isentropic perturbation equation,} or HAIPE:
\begin{myEquationFrame}[frametitle = \colorbox{white}{HAIPE}]
    % \begin{equation}
    %     \begin{split}
    %         \tfrac{1}{3}\kappa(1-3c_s^2)\tensor{q}{_\alpha}\tensor{\delta u}{^\alpha}&=\ddot{\psi}-c_s^2\lozenge\psi+\tfrac{1}{3}\tensor{\dot{u}}{^\alpha}\tensor{\psi}{_{;\alpha}}+\tfrac{4}{3}\theta\dot{\psi} \\
    %         &\qquad+\left[\tfrac{4}{3}\dot{\theta}+\tfrac{2}{3}(1+c_s^2)\theta^2+2(1-c_s^2)\sigma^2+\tfrac{2}{3}(1-9c_s^2)\omega^2-c_s^2(R+2\tensor{R}{_\alpha_\beta}\tensor{u}{^\alpha}\tensor{u}{^\beta})\right]\psi
    %     \end{split}
    % \end{equation}
    \begin{equation}\label{eq:masterEquations}
        \begin{split}
            &\ddot{\psi}-c_s^2\lozenge\psi+\tfrac{1}{3}\tensor{\dot{u}}{^\alpha}\tensor{\psi}{_{;\alpha}}+\tfrac{4}{3}\theta\dot{\psi}+\left(\tfrac{4}{3}\dot{\theta}+(1+c_s^2)(\tfrac{2}{3}\theta^2-\Lambda)+2(1-c_s^2)\sigma^2+\tfrac{2}{3}(1-9c_s^2)\omega^2\right)\psi \\ 
            &\qquad=2\kappa(1-3c_s^2)\,\tensor{q}{_\alpha}\tensor{\pertnb u}{^\alpha}.
        \end{split}
    \end{equation}
\end{myEquationFrame}
This is the equation that governs isentropic (so also adiabatic) evolutions of perturbations, in the Newtonian gauge. One notices that the rhs of this equation now depends only $\tensor{q}{^\mu}$. Hence, if that vector vanishes (and we shall see in the next section it usually does), the rhs vanishes and we have a damped but undriven wave equation in $\psi$.

\subsection{Tensor perturbations}
Gravitational waves are essentially tensor perturbations in spacetime. We can also investigate how these would travel in such anisotropic spacetimes. Concretely, thus, we assume a perturbation of the form
\begin{equation}
    \tensor{\pertnb g}{_\mu_\nu}=2\tensor{E}{_\mu_\nu},
\end{equation}
where $\boldsymbol{E}$ has the properties that
\begin{equation}
    \tensor{E}{_\mu_\nu}=\tensor{E}{_{\langle\mu\nu\rangle}},\quad\tensor{E}{^\mu^\alpha_{|\alpha}}\equiv\tensor*{h}{^\mu_\alpha}\tensor*{h}{^\gamma_\beta}\tensor{E}{^\alpha^\beta_{;\gamma}}=\tb{0}.
\end{equation}
These properties imply that we have the following identities at our disposal:
\begin{subequations}
    \begin{align}
        \tensor{u}{_\alpha}\tensor{h}{_\beta_\gamma}\tensor{E}{^\alpha^\beta^{;\gamma}}&=-\tensor{\sigma}{_\alpha_\beta}\tensor{E}{^\alpha^\beta} \\
        \tensor{\dot{u}}{_\alpha}\tensor{h}{_\beta_\gamma}\tensor{E}{^\alpha^\beta^{;\gamma}}&=0 \\
        \tensor{E}{^\alpha^\beta_{;\alpha\beta}}&=\tensor{\sigma}{_\alpha_\beta}\tensor{\dot{E}}{^\alpha^\beta}+\tensor{E}{^\alpha^\beta}(\tensor{\dot{\sigma}}{_\alpha_\beta}+\theta\,\tensor{\sigma}{_\alpha_\beta}+\tensor{\dot{u}}{_\alpha}\tensor{\dot{u}}{_\beta}+\tensor{\dot{u}}{_{\alpha|\beta}}) \\
        \tensor{u}{_\beta}\tensor{u}{_\gamma}\tensor{E}{^\alpha^\beta^{;\gamma}_{;\alpha}}&=-\tensor{\sigma}{_\alpha_\beta}\tensor{\dot{E}}{^\alpha^\beta}+\tensor{E}{^\alpha^\beta}(\tensor{u}{_\alpha^{|\gamma}}\tensor{u}{_\gamma_{|\beta}}-\tensor{\dot{u}}{_\alpha_{|\beta}}-2\,\tensor{\dot{u}}{_\alpha}\tensor{\dot{u}}{_\beta}) \\
        \tensor{u}{_\alpha}\tensor{u}{_\beta}\,\square\tensor{E}{^\alpha^\beta}&=2\,\tensor{E}{^\alpha^\beta}(\tensor{u}{_\alpha^{|\gamma}}\tensor{u}{_\beta_{|\gamma}}-\tensor{\dot{u}}{_\alpha}\tensor{\dot{u}}{_\beta}).
    \end{align}
\end{subequations}
In a similar fashion as we did in \S\ref{sec:scalarPerts}, we can now evaluate the perturbation equations for $\tb{E}$. For this, we once more utilize the \texttt{xPand} package for Mathematica \cite{pitrou_xpand_2013}. After manual simplification, we obtain the following components of the perturbed Einstein tensor:
\begin{subequations}
    \begin{align}
        \tensor{u}{^\alpha}\tensor{u}{^\beta}\tensor{\pertnb G}{_\alpha_\beta}&=-\tensor{\sigma}{_\alpha_\beta}\tensor{\dot{E}}{^\alpha^\beta}+\tensor{E}{^\alpha^\beta}(-\tensor{R}{_\alpha_\beta}+4\,\tensor{u}{_\alpha^{|\gamma}}\tensor{\omega}{_\gamma_\beta}+\tensor{\dot{\sigma}}{_\alpha_\beta}+\theta\,\tensor{\sigma}{_\alpha_\beta}-\tensor{\dot{u}}{_\alpha_{|\beta}}-\tensor{\dot{u}}{_\alpha}\tensor{\dot{u}}{_\beta}) \\
        \tensor{h}{^\alpha^\beta}\tensor{\pertnb G}{_\alpha_\beta}&=-3\,\tensor{\sigma}{_\alpha_\beta}\tensor{\dot{E}}{^\alpha^\beta}+\tensor{E}{^\alpha^\beta}(3\,\tensor{R}{_\alpha_\beta}+4\,\tensor{u}{_\alpha^{|\gamma}}\tensor{\omega}{_\gamma_\beta}-3\,\tensor{\dot{u}}{_\alpha_{|\beta}}-3\,\tensor{\dot{u}}{_\alpha}\tensor{\dot{u}}{_\beta}-\tensor{\dot{\sigma}}{_\alpha_\beta}-\theta\,\tensor{\sigma}{_\alpha_\beta}) \\
        \tensor{u}{^\alpha}\tensor*{h}{_\mu^\beta}\tensor{\pertnb G}{_\alpha_\beta}&= \tensor{u}{^\alpha}\tensor*{h}{^\beta_\mu}\left(2\,\tensor*{E}{^\gamma_{(\alpha;\beta)}_{;\gamma}}-\square\tensor{E}{_\alpha_\beta}\right)\\
        \tensor{\pertnb G}{_{\langle\mu\nu\rangle}}&=-R\,\tensor{E}{_\mu_\nu}+2\,\tensor*{E}{^\alpha_{\langle\mu;\nu\rangle}_\beta}-\square\tensor{E}{_{\langle\mu\nu\rangle}}
    \end{align}
\end{subequations}
In the final equation, the PSTF step $\langle\,\dots\rangle$ is supposed to act \emph{after} all the differentials have been applied. We note that particularly the first two equation simplify considerably if $\tb{\sigma}=\tb{\omega}=\tb{\dot{u}}=0$, like in the standard FLRW case. Using the perturbed Einstein equation and \eqref{eq:EMTExpansions}, we can thus find the following decomposed perturbation equations.
\begin{subequations}\label{eq:tensor-perts}
    \begin{align}
        \kappa\,\pertnb\rho&=\tensor{\sigma}{_\alpha_\beta}\tensor{\dot{E}}{^\alpha^\beta}+\tensor{E}{^\alpha^\beta}(-\tensor{R}{_\alpha_\beta}+4\,\tensor{u}{_\alpha^{|\gamma}}\tensor{\omega}{_\gamma_\beta}+\tensor{\dot{\sigma}}{_\alpha_\beta}+\theta\,\tensor{\sigma}{_\alpha_\beta}-\tensor{\dot{u}}{_\alpha_{|\beta}}-\tensor{\dot{u}}{_\alpha}\tensor{\dot{u}}{_\beta})-2\kappa\,\tensor{q}{_\alpha}\tensor{\pertnb u}{^\alpha} \label{eq:gravwav-1}\\
        \begin{split}
            3\kappa\,\pertnb p&=3\,\tensor{\sigma}{_\alpha_\beta}\tensor{\dot{E}}{^\alpha^\beta}+\tensor{E}{^\alpha^\beta}(3\,\tensor{R}{_\alpha_\beta}+4\,\tensor{u}{_\alpha^{|\gamma}}\tensor{\omega}{_\gamma_\beta}-3\,\tensor{\dot{u}}{_\alpha_{|\beta}}-3\,\tensor{\dot{u}}{_\alpha}\tensor{\dot{u}}{_\beta}-\tensor{\dot{\sigma}}{_\alpha_\beta}-\theta\,\tensor{\sigma}{_\alpha_\beta}) \\
            &\qquad-2\kappa\,\tensor{\pi}{_\alpha_\beta}\tensor{E}{^\alpha^\beta}-2\kappa\,\tensor{q}{^\alpha}\tensor{\pertnb u}{_\alpha}
        \end{split} \label{eq:gravwav-2} \\
        \kappa\,\tensor{\pertnb q}{^{\langle\mu\rangle}}&=-\tensor{u}{^\alpha}\tensor*{h}{^\beta_\mu}\left(2\,\tensor*{E}{^\gamma_{(\alpha;\beta)}_{;\gamma}}-\square\tensor{E}{_\alpha_\beta}\right)-\kappa\left(((\rho+p)\tensor*{h}{^\mu_\alpha}+\tensor*{\pi}{_\alpha^\mu})\tensor{\pertnb u}{^\alpha}+2\,\tensor{q}{_\alpha}\tensor{E}{^\mu^\alpha}\right) \\
        \kappa\,\tensor{\pertnb\pi}{_{\langle\mu\nu\rangle}}&=-R\,\tensor{E}{_\mu_\nu}+2\,\tensor*{E}{^\alpha_{\langle\mu;\nu\rangle}_\beta}-\square\tensor{E}{_{\langle\mu\nu\rangle}}-\kappa\,\tensor{q}{_{\langle\mu}}\tensor{\pertnb u}{_{\nu\rangle}}.
    \end{align}
\end{subequations}
These are thus the perturbation equations for gravitational waves. As a final step, we can with \eqref{eq:soundwaves} and under the assumption of isentropy relate \eqref{eq:gravwav-1} and \eqref{eq:gravwav-2} into one equation:
\begin{equation}\label{eq:one_tensor_pert}
    \begin{split}
        2\kappa(1-3c_s^2)\,\tensor{q}{_\alpha}\tensor{\pertnb u}{^\alpha}&=3(1-c_s^2)\,\tensor{\sigma}{_\alpha_\beta}\tensor{\dot{E}}{^\alpha^\beta}+\tensor{E}{^\alpha^\beta}\left(3(1+c_s^2)\,\tensor{R}{_\alpha_\beta}+4(1-3c_s^2)\tensor{u}{_\alpha^{|\gamma}}\tensor{\omega}{_\gamma_\beta}\right. \\
        &\qquad\left.-(1+3c_s^2)(\tensor{\dot{\sigma}}{_\alpha_\beta}+\theta\,\tensor{\sigma}{_\alpha_\beta})+3(c_s^2-1)(\tensor{\dot{u}}{_\alpha_{|\beta}}+\tensor{\dot{u}}{_\alpha}\tensor{\dot{u}}{_\beta})-2\kappa\,\tensor{\pi}{_\alpha_\beta}\right)
    \end{split}
\end{equation}
As above, in circumstances where $\tb{q}\equiv0$, the entire lhs vanishes identically. Although this equation may prove useful in some specialized circumstances, in and of itself it cannot determine the entire behavior of the perturbation (e.g. when given some initial conditions), for the reason that it is merely one equation, and $\tb{E}$ has more than one degree of freedom.

\section{A specific anisotropic metric choice}\label{sec:metricChoice}
The result of the previous two subsections was the finding of equations \eqref{eq:masterEquations}, \eqref{eq:tensor-perts}, and \eqref{eq:one_tensor_pert}. Of these, the former is the equation governing isentropic scalar perturbations, when assuming $\phi=\psi$ in the perturbation \eqref{eq:pert-newton-gauge}, and having chosen Newtonian gauge. This is thus the generalization of the Mukhanov-Sasaki perturbation equation, which is applicable only to conformally flat FLRW spacetimes with $\tensor{u}{^\mu}=a^{-1}\tensor*{\delta}{^\mu_0}$. In contrast, using \eqref{eq:masterEquations}, we can investigate any spatially homogeneous spacetime, and with any fluid flow vector that we wish, so long as it has been normalized. The aim of this subsection is to calculate the terms featuring in the equations \eqref{eq:masterEquations}, \eqref{eq:tensor-perts}, and \eqref{eq:one_tensor_pert}, featuring a common choice of metric components and fluid flow that nevertheless simplifies the calculations considerably.

Since we wish to consider spatially homogeneous universes, we shall choose to work in an arbitrary, i.e.\ generally non-coordinated, frame: as explained in Section \ref{sec:homgeom}, this is the most natural way in which to approach this situation. Let $\tb{g}$ be the metric of the spatially homogeneous spacetime we wish to investigate, and let us make the assumptions that (i) the metric, when expanded in the frame as introduced in Section \ref{sec:homgeom}, is diagonal, and (ii) the fluid flow is $\tensor{u}{^\mu}=\tensor*{\delta}{^\mu_0}$, in that same frame. Writing $\{\tensor{\tb{X}}{_\mu}\}_{\mu=0}^3$ and $\{\tensor{\tb{e}}{^\mu}\}_{\mu=0}^3$ for the frame and coframe, respectively, we thus have
\begin{equation}
    \tb{g}=\tensor{g}{_\mu_\nu}(t)\,\tensor{\tb{e}}{^\mu}\tensor{\tb{e}}{^\nu},\qq{where}\tensor{g}{_\mu_\nu}(t)=\operatorname{diag}(-1,a_1(t)^2,a_2(t)^2,a_3(t)^2),
\end{equation}
where $t$ is the cosmic time featuring as $\dd{t}\equiv\tensor{\tb{e}}{^0}$, and $\tb{u}=\tensor{\tb{X}}{_0}$. As a consequence of the exactness of $\tensor{\tb{e}}{^0}$, we have the following commutation relations for the frame:
\begin{equation}
    [\tensor{\tb{X}}{_0},\tensor{\tb{X}}{_i}]=\tensor{C}{^a_0_i}\tensor{\tb{X}}{_a}\qand [\tensor{\tb{X}}{_i},\tensor{\tb{X}}{_j}]=\tensor{C}{^a_i_j}\tensor{\tb{X}}{_a},
\end{equation}
where the $\tb{C}$ feature as the structure constants of the frame. In fact, for our situation we have that $\tensor{C}{^i_0_j}=0$ for all $i,j$, as we wish to have the frame be independent of time.

We have that $\tensor{u}{_\alpha}\tensor{u}{^\alpha}=-1$, so our fluid flow is normalized; and additionally we have $\tensor{\dot{u}}{^\mu}=\tensor{\Gamma}{^\mu_\alpha_\beta}\tensor{u}{^\alpha}\tensor{u}{^\beta}=0$. That is, our fluid flow is geodesic in this metric, even independently of what our structure constants are. We would expect this physically: since the fluid flow is merely in the temporal direction, and the non-coordinated facet of our situation appears in the spatial directions, there should be no influence. The projection tensor is given by
\begin{equation}
    \tensor{h}{_\mu_\nu}=\operatorname{diag}(0,a_1^2,a_2^2,a_3^2).
\end{equation}

In order to populate \eqref{eq:masterEquations} given this metric and fluid flow, we shall need to calculate the quantities listed below.
\begin{description}
    \item[The kinematical tensors] We calculate these based on their definitions in \ref{sec:kinematicalTensors}.
    \begin{description}
        \item[\textit{The expansion scalar}] By definition, $\theta:=\tensor{u}{^\alpha_{;\alpha}}$. Straightforwardly,
        \begin{equation}
            \theta=\tensor{u}{^\alpha_{;\alpha}}=\tensor{\Gamma}{^\alpha_0_\alpha}=\sum_{i=1}^3H_i\qand\dot{\theta}=\sum_{i=1}^3\dot{H_i}.
        \end{equation}

        \item[\textit{The shear tensor \& scalar}] The definition is $\tensor{\sigma}{_\mu_\nu}=\tensor{u}{_{\langle\mu;\nu\rangle}}$. As such, the shear tensor and scalar are given by
        \begin{equation}
            \tensor{\sigma}{_\mu_\nu}=\tensor*{h}{^{(\alpha}_\mu}\tensor*{h}{^{\beta)}_\nu}\tensor{u}{_\alpha_{;\beta}}-\tfrac{1}{3}\theta\,\tensor{h}{_\mu_\nu}=\begin{cases}a_i\dot{a_i}-\tfrac{1}{3}\theta a_i^2 & \qif\mu=\nu=i \\ 0 &\qotherwise\end{cases}
        \end{equation}
        and
        \begin{equation}
            2\sigma^2=\tensor{\sigma}{^\alpha^\beta}\tensor{\sigma}{_\alpha_\beta}=\sum_{i=1}^3(a_i^{-3}\dot{a_i}-\tfrac{1}{3}\theta a_i^{-2})(a_i\dot{a_i}-\tfrac{1}{3}\theta a_i^2)=\sum_{i=1}^3(H_i-\tfrac{1}{3}\theta)^2.
        \end{equation}

        \item[\textit{The vorticity tensor \& scalar}] The definition is $\tensor{\omega}{_\mu_\nu}=\tensor{u}{_{[\mu|\nu]}}$. Since $\tensor{u}{_{\mu|\nu}}=\tensor{u}{_\mu_{;\nu}}=\tensor{\Gamma}{^0_\mu_\nu}$ is symmetric, the vorticity vanishes.
    \end{description}

    \item[The projected wave operator] The definition of the projected wave operator acting on a scalar function $f$ (or a tensorial quantity) is $\lozenge f:=\tensor{h}{^\mu^\nu}\tensor{f}{_{;\mu\nu}}$ cf. Definition \ref{def:pwo}. Knowing the metric and the connection, we can write this in the more explicit form
    \begin{equation}
        \lozenge f=\tensor{h}{^\mu^\nu}(\tensor{f}{_{,\mu\nu}}-\tensor{\Gamma}{^\alpha_\mu_\nu}\tensor{f}{_{,\alpha}})=\left[\sum_{i=1}^3a_i^{-2}\left(\tensor{\tb{X}}{_i}\tensor{\tb{X}}{_i}-\tensor{\Gamma}{^\alpha_i_i}\,\tensor{\tb{X}}{_\alpha}\right)\right]f.
    \end{equation}
    For completeness, we also note that $\ddot{f}=\tensor{\tb{X}}{_0}\tensor{\tb{X}}{_0}f$.

    \item[The Einstein tensor] We notice that we are in the case as outlined in Wald \cite[\S7.2]{wald_general_1984}, where the components of the Einstein tensor are presented---we shall thus use those expressions for the components. We first calculate
    \begin{equation}
        \tensor{K}{_\mu_\nu}:=\tfrac{1}{2}\mathcal{L}_{\vb*{u}}\tensor{h}{_\mu_\nu}=\tfrac{1}{2}\tensor{h}{_\mu_\nu_{;\alpha}}\tensor{u}{^\alpha}+\tensor{h}{_\alpha_{(\mu}}\tensor{u}{^\alpha_{;\nu)}}=\tensor{\dot{u}}{_{(\mu}}\tensor{u}{_{\nu)}}+\tensor{u}{_{(\mu;\nu)}}=\tensor{u}{_{(\mu|\nu)}}=\tfrac{1}{3}\theta\tensor{h}{_\mu_\nu}+\tensor{\sigma}{_\mu_\nu},
    \end{equation}
    so that we additionally have $(\tensor*{K}{^\alpha_\alpha})^2=\theta^2$ and $\tensor{K}{^\alpha^\beta}\tensor{K}{_\alpha_\beta}=\tfrac{1}{3}\theta^2+2\sigma^2$. We can calculate moreover
    \begin{equation}
        \mathcal{L}_{\vb*{u}}\tensor{K}{_\mu_\nu}=\tensor{\dot{\sigma}}{_\mu_\nu}+\tfrac{1}{3}(\dot{\theta}+\tfrac{2}{3}\theta^2)\tensor{h}{_\mu_\nu}+\tfrac{2}{3}\theta\tensor{\sigma}{_\mu_\nu}+2\tensor{\sigma}{_\alpha_{(\mu}}\tensor{u}{^\alpha_{;\nu)}}.
    \end{equation}
    Thus, we obtain the following components of the Einstein tensor:
    \begin{subequations}
        \begin{align}
            \tensor{u}{^\alpha}\tensor{u}{^\beta}\tensor{G}{_\alpha_\beta}&=\tfrac{1}{3}\theta^2-\sigma^2-\tfrac{1}{2}\tensor{C}{^\alpha_\alpha_\beta}\tensor{C}{_\gamma^\gamma^\beta}-\tfrac{1}{4}\tensor{C}{^\alpha_\gamma_\beta}\tensor{C}{^\gamma_\alpha^\beta}-\tfrac{1}{8}\tensor{C}{_\alpha_\beta_\gamma}\tensor{C}{^\alpha^\beta^\gamma}, \label{eq:friedmann_rho}\\
            \tensor{h}{^\alpha^\beta}\tensor{G}{_\alpha_\beta}&=-2\dot{\theta}-\theta^2-3\sigma^2+\tfrac{1}{2}\tensor{C}{^\alpha_\alpha^\beta}\tensor{C}{^\gamma_\gamma_\beta}+\tfrac{1}{4}\tensor{C}{^\alpha_\beta_\gamma}\tensor{C}{^\beta_\alpha^\gamma}+\tfrac{1}{8}\tensor{C}{_\alpha_\beta_\gamma}\tensor{C}{^\alpha^\beta^\gamma}, \label{eq:friedmann_p}\\
            \tensor{h}{^\mu^\alpha}\tensor{u}{^\beta}\tensor{G}{_\alpha_\beta}&=\tensor{\sigma}{^\mu^\alpha}\tensor{C}{^\beta_\alpha_\beta}+\tensor*{\sigma}{^\alpha_\beta}\tensor{C}{^\beta_\alpha^\mu}, \\
            \begin{split}
                \tensor{G}{_{\langle\mu\nu\rangle}}&=\tensor{\dot{\sigma}}{_\mu_\nu}+\theta\tensor{\sigma}{_\mu_\nu}+\tfrac{1}{8}\tensor{C}{_\mu_\alpha_\beta}\tensor{C}{_\nu^\alpha^\beta}+\tensor{C}{_\nu_\alpha_\beta}\tensor{C}{_\mu^\alpha^\beta}-\tensor{C}{^\alpha_\alpha_\beta}\tensor{C}{_{\langle\mu\nu\rangle}^\beta}-\tensor{C}{_\alpha_\beta_{\langle\mu}}\tensor{C}{^{(\alpha\beta)}_{\nu\rangle}}\\
                &\qquad-\tfrac{1}{12}\tensor{C}{_\alpha_\beta_\gamma}\tensor{C}{^\alpha^\beta^\gamma}\,\tensor{h}{_\mu_\nu}.
            \end{split} \label{eq:friedmann_pi}
        \end{align}
    \end{subequations}
    We also used that, through the Raychaudhuri equation, the Ricci scalar is found to be
    \begin{equation}
        R=2\,\tensor{u}{^\alpha}\tensor{u}{^\beta}(\tensor{G}{_\alpha_\beta}-\tensor{R}{_\alpha_\beta})=2\dot{\theta}+\tfrac{4}{3}\theta^2+2\sigma^2-\tensor{C}{^\alpha_\alpha_\beta}\tensor{C}{_\gamma^\gamma^\beta}-\tfrac{1}{2}\tensor{C}{^\alpha_\gamma_\beta}\tensor{C}{^\gamma_\alpha^\beta}-\tfrac{1}{4}\tensor{C}{_\alpha_\beta_\gamma}\tensor{C}{^\alpha^\beta^\gamma}.
    \end{equation}
    The term involving the structure constants in \eqref{eq:friedmann_rho} and \eqref{eq:friedmann_p} may be interpreted as (positive resp.\ negative) one half of the 3-scalar curvature of the hypersurface, in the sense of the Gauss equation governing the curvature of an embedded surface. \cite{maccallum_cosmological_1973}. Similarly, the term involving the structure constants in \eqref{eq:friedmann_pi} is the traceless part of the 3-Ricci tensor.

    \item[The Friedmann equations]
    Utilizing the found components of the Einstein tensor, we can fill in the Einstein equation and find the following Friedmann equations for our cosmology.
    \begin{subequations}
        \begin{align}
            \kappa\,\rho&=\tfrac{1}{3}\theta^2-\sigma^2-\Lambda-\tfrac{1}{2}\tensor{C}{^\alpha_\alpha_\beta}\tensor{C}{_\gamma^\gamma^\beta}-\tfrac{1}{4}\tensor{C}{^\alpha_\gamma_\beta}\tensor{C}{^\gamma_\alpha^\beta}-\tfrac{1}{8}\tensor{C}{_\alpha_\beta_\gamma}\tensor{C}{^\alpha^\beta^\gamma}, \\
            3\kappa\,p&=-2\dot{\theta}-\theta^2-3\sigma^2+3\Lambda+\tfrac{1}{2}\tensor{C}{^\alpha_\alpha^\beta}\tensor{C}{^\gamma_\gamma_\beta}+\tfrac{1}{4}\tensor{C}{^\alpha_\beta_\gamma}\tensor{C}{^\beta_\alpha^\gamma}+\tfrac{1}{8}\tensor{C}{_\alpha_\beta_\gamma}\tensor{C}{^\alpha^\beta^\gamma}, \\
            \kappa\,\tensor{q}{^\mu}&=-\tensor{\sigma}{^\mu^\alpha}\tensor{C}{^\beta_\alpha_\beta}-\tensor*{\sigma}{^\alpha_\beta}\tensor{C}{^\beta_\alpha^\mu}, \\
            \begin{split}
                \kappa\,\tensor{\pi}{_\mu_\nu}&=\tensor{\dot{\sigma}}{_\mu_\nu}+\theta\tensor{\sigma}{_\mu_\nu}+\tfrac{1}{8}\tensor{C}{_\mu_\alpha_\beta}\tensor{C}{_\nu^\alpha^\beta}+\tensor{C}{_\nu_\alpha_\beta}\tensor{C}{_\mu^\alpha^\beta}-\tensor{C}{^\alpha_\alpha_\beta}\tensor{C}{_{\langle\mu\nu\rangle}^\beta}-\tensor{C}{_\alpha_\beta_{\langle\mu}}\tensor{C}{^{(\alpha\beta)}_{\nu\rangle}}\\
                &\qquad-\tfrac{1}{12}\tensor{C}{_\alpha_\beta_\gamma}\tensor{C}{^\alpha^\beta^\gamma}\,\tensor{h}{_\mu_\nu}.
            \end{split}
        \end{align}
    \end{subequations}
    Noteworthy in these equations is that all the terms involving the structure constants have, in sum, one of their indices raised w.r.t.\ the definitions. This means that these contributions scale as $\propto a_i^{-2}$, for the appropriate scale factor $a_i$, so the effect/strength is similar to the influence of curvature as known from the open and closed FLRW universes. Furthermore, we note that in order to effect non-zero momentum density $\tb{q}$ and anisotropic stress tensor $\tb{\pi}$, necessary conditions are (i) non-zero shear, so no identical scale factors, and (ii) non-zero structure constants, so no coordinate frame.
 \end{description}

\section{Example: density contrasts}\label{sec:density_contrasts}
To assess the validity of the obtained expressions, we first investigate its results in the simple situation of a perfectly homogeneous and isotropic Universe, i.e.\ a conventional FLRW cosmology with a Robertson-Walker metric, and that for the anisotropic Bianchi I model. To this end, we define the mass \emph{density contrast} $\delta(x)$,
\begin{equation}
    \delta(x):=\frac{\rho(x)-\overline{\rho}(t)}{\overline{\rho}(t)}\iff \rho=(1+\delta)\overline{\rho}\,.
\end{equation}
with $\rho(x)$ the (energy) density at some spacetime event $x$ and global energy density $\overline{\rho}(t)$ given by some cosmological model \footnote{Note that a choice of gauge here is employed to make this relation.}. It implies the equivalent definition for $\delta:=\pertnb\rho/\rho$. Qualitatively $\delta>0$ means that there is a local excess of energy, whereas $\delta<0$ means there is a deficit, with a physical lower bound of $\delta\geq-1$.

In the situation of a FLRW cosmology, the obtained expressions should yield an evolution of the density contrast evolution that, in the early linear regime,is in accordance with the predictions of the (Newtonian) theory of perturbation growth in FLRW cosmologies (see \cite{peebles_large-scale_1980}).

Knowing the forms of $\pertnb\rho$ and $\rho$ (from previous sections), on the basis of the inferred expression for the scalar perturbations (\S\ref{sec:scalarPerts}) we may (i) infer the explicit expression of the density contrast, and (ii) make qualitative predictions concerning this. For the stated purposes, we consider two explicit circumstances. These are those of (i) the flat FLRW Einstein-de Sitter (EdS) case, and (ii) a Bianchi I universe. The EdS situation serves as a testbed for confirming the validity of our inferred perturbation equations. Note that our methodology \emph{does not limit which type of Bianchi model we can analyze:} it is only for computational ease that we include Bianchi I in the present study.

\subsection{Einstein-de Sitter universe}
The flat FLRW EdS universe translates to our case---in the notation of \S\ref{sec:metricChoice}---with the following parameter choices:
\begin{equation}
    \tb{C}\equiv\tb{0},\quad a_1(t)=a_2(t)=a_3(t)\equiv a(t),\qand c_s=0.
\end{equation}
The relations for the corresponding isotropic flow field and expansion are $\theta=3\,\dot{a}/a$ and
$\tb{q}=\tb{\sigma}=\tb{\pi}=\tb{0}$. This immediately yields the non-trivial Friedmann equations,
\begin{equation}
  \kappa\rho=\tfrac{1}{3}\theta^2-\Lambda,\qand
  -2\dot{\theta}-\theta^2+3\Lambda=0,
\end{equation}
which implies the following expression for the evolving density perturbation $\pertnb\rho$ and
\emph{homogeneous \& anisotropic, isentropic perturbation equation,} (HAIPE)
\begin{equation}
    \kappa\,\pertnb\rho=2\lozenge\psi-\tfrac{2}{3}\theta^2\psi,\qand
    3\ddot{\psi}+4\theta\dot{\psi}+[4\dot{\theta}+2\theta^2-3\Lambda]\psi=0\,.
\end{equation}

For an Einstein-de Sitter Universe, we may directly write the explicit dependence of expansion factor $a$ and
expansion rate $\theta$ on cosmic time $t$. It also involves a zero cosmological constant $\Lambda=0$. Inserting these, along with the second Friedmann equation, and simplifying the HAIPE yields 
\begin{equation}
    3\,\tensor{\psi}{_{,00}}+8t^{-1}\,\tensor{\psi}{_{,0}}=0\implies\psi\propto t^{0}\qor \psi\propto t^{-5/3}.
\end{equation}
From this expression, we obtain the expression for the time evolution of the density contrast, 
\begin{equation}\label{eq:edsperturb}
  \tfrac{1}{2}\delta=\frac{\kappa\,\pertnb\rho}{2\,\kappa\rho}=\frac{\lozenge{\psi}}{\frac{1}{3}\theta^2}-\psi=\frac{3}{a^2\theta^2}\,\tensor{\delta}{^a^b}\tensor{\psi}{_{,ab}}-\frac{3}{a^2\theta}\tensor{\psi}{_{,0}}-\psi.
\end{equation}
Inserting the time evolutions for $a$, $\theta$, and $\psi$, we find that the first term in this expression evolves as $t^{2/3}$ and $t^{-1}$, the second term as $t^{-2}$, and the last term as $t^0$ and $t^{-5/3}$. For the time evolution of the density contrast, we would then conclude that the leading terms would be $t^{2/3}$ and constancy in time. This stands in contrast to the generally established result for Newtonian perturbation theory, which yields a density contrast evolution $\delta\propto t^{2/3},t^{-1}$ \cite[Eq.~(10.123)]{peebles_principles_1993}.

However, when inspecting equation \eqref{eq:edsperturb} we may notice that for linear circumstances the amplitude
of $\psi$ is very small. Ignoring the corresponding term in equation \eqref{eq:edsperturb}, we would find the
that the density contrast would evolve the same as in Newtonian perturbation theory. It provides confidence
in the scalar, and also tensor, perturbation equations that we inferred.

\subsection{Bianchi I universe}
Subsequently, we turn to the situation of a Bianchi I universe. We address the configuration of a
Bianchi I universe without pressure, i.e.\ ``cosmological dust'', and without a cosmological constant, i.e.\ $\Lambda=0$. It involves the choice $\tb{C}\equiv\tb{0}$, while a pressureless medium, i.e.\ $p=0$,  also implies a zero sound velocity, $c_s=0$. Consequently, we have $\tb{q}=\tb{0}$, and the non-trivial Friedmann equations
\begin{equation}
  \kappa\rho=\tfrac{1}{3}\theta^2-\sigma^2,\quad2\dot{\theta}+\theta^2+3\sigma^2=0,\qand\kappa\,\tensor{\pi}{_\mu_\nu}=\tensor{\dot{\sigma}}{_\mu_\nu}+\theta\,\tensor{\sigma}{_\mu_\nu}\,.
\end{equation}
The \emph{homogeneous \& anisotropic, isentropic perturbation equation} (HAIPE), and density perturbation
equation are then:
\begin{equation}
  \ddot{\psi}+\tfrac{4}{3}\theta\dot{\psi}+2\psi\left(\tfrac{2}{3}\dot{\theta}+\tfrac{1}{3}\theta^2+\sigma^2\right)=0,\qand\kappa\,\pertnb\rho=2\lozenge\psi+2\psi\left(\sigma^2-\tfrac{1}{3}\theta^2\right).
\end{equation}
We see that the HAIPE reduces to the case also studied above. Hence, the implied time evolutions are also 
power laws, with $\psi\sim t^0$ or $\psi\sim t^{-5/3}$. The corresponding density contrast is
\begin{equation}
    \tfrac{1}{2}\delta=\frac{\lozenge\psi+\psi(\sigma^2-\tfrac{1}{3}\theta^2)}{\tfrac{1}{3}\theta^2-\sigma^2}=3\theta^{-2}\lozenge\psi-\psi+\left(\frac{\sigma}{\theta}\right)^{\!2}\times 9\theta^{-2}\lozenge\psi+\order{(\sigma/\theta)^4},
\end{equation}
where we assumed that the ratio $\sigma/\theta\ll1$ in order to leverage a geometric series expansion. Assuming as before that the addition of $\psi$ can be disregarded, we see that the presence of shear tends to exacerbate any existing density contrast there is. That is to say, if $\delta|_{\sigma=0}>0$, then $\delta|_{\sigma>0}>\delta|_{\sigma=0}$, and vice versa for negative values of $\delta|_{\sigma=0}$.

\medskip
As an example of this in action, consider the following rough example of an overdensity $\delta>0$ of a domain contained in a slice of constant cosmic time time. In the case of vanishing shear, $\sigma=0\iff\tb{\sigma}=\tb{0}$, a local volume element is isotropic, and thus it exhibits spherical symmetry. This implies that its surface area is at a minimum for the amount of volume that it contains, and thus that any exchange of energy with the environment is also minimized. In contrast, for non-vanishing shear the relevant surface area will necessarily be greater than the case of vanishing shear, facilitating greater exchanges the environment. Since the rate of exchange with the environment in the case of an overdensity means accumulation of more energy, an overdensity in non-vanishing shear will accumulate more rapidly than for vanishing shear. Under the additional assumption of vanishing anisotropic pressure, this contribution will tend to lessen for greater cosmic time. Indeed, it is known that in the absence of a cosmological constant, we will have that $\sigma/\theta\to0$ for the Bianchi I class of models \cite[\S18.2]{ellis_relativistic_2012} (and in fact this holds for a broader class of Bianchi models, see \cite{wald_asymptotic_1983}).

\section{Discussion and conclusion}
Based on an existing and growing interest in anisotropic alternatives to the standard FLRW cosmologies, in this article we treated perturbations to generic Bianchi cosmological models. We saw that if we express the metric in a frame that is naturally (or by construction) adapted to them (and not necessarily coordinated), the metric components will depend only on time, and the vector fields defining the frame will define a Lie algebra. This allows us to consider General Relativity as if we were working in a coordinate frame, but for some additions to the Christoffel symbols of the Levi-Civita connection and the Riemann tensor.

After briefly introducing perturbation theory from a mathematical point of view, we proceeded to perturb the metric, energy-momentum, and Einstein tensors to the first order. Then, by fixing the Newtonian gauge, we investigated pure scalar and pure tensor perturbations, deriving for both their effects on the quantities in the energy-momentum tensor, i.e.\ on the energy density $\rho$, relativistic pressure $p$, momentum density $\tb{q}$, and anisotropic stress tensor $\tb{\pi}$. In the former case, under suitable assumptions, we combined the density and pressure equations in order to obtain a ``master'' perturbation equation \eqref{eq:masterEquations}, comparable to the Mukhanov-Sasaki equation known from this analysis in the FLRW case, and found in e.g. \cite[\S10.2.9]{ellis_relativistic_2012}.

Finally, we fixed the metric components to be diagonal and anisotropic, $\tensor{g}{_\mu_\nu}=\operatorname{diag}(-1,a_1(t)^2,a_2(t)^2,a_3(t)^2)$, and the fluid flow spatially stationary, $\tensor{u}{^\mu}=\tensor*{\delta}{^\mu_0}$, deriving for this choice expressions of the kinematical quantities. These were necessary in order to fill in the general perturbation equations derived earlier. Additionally, based on Wald \cite{wald_general_1984}, we derived the Friedmann equations for this universe, which featured to a great degree the structure constants of the frame's Lie algebra. This information populates the perturbation equation(s), which allows them to be solved.

%An issue with this analysis, and that we also discussed in \S\ref{sec:gaugeTransfos}, is that it was necessary to fix the gauge in order to progress further with the perturbation results. Since our focus is on structure formation and CMB analyses, and fixing (Newtonian) gauge is implicitly utilized there, this does not pose a major problem. Our analysis admits generalization if the (technical) difficulties with general gauge transformations are resolved. This would then pave the way for generalized scalar, tensor, and vector perturbation equations in anisotropic universes, together with the appropriate gauge transformation rules between different choices of gauge.

Though we chose in \S\ref{sec:metricChoice} to study quite a general class of metric (diagonal with distinct scale factors, and stationary flow), this is not the most general situation that can be imagined---particularly with regards to the fluid flow $\tb{u}$. Namely, the property of our fortuitous choice that does the heavy lifting is that \emph{the fluid flow is geodesic in this metric,} regardless of choice of structure constants. Although $\tensor{u}{^\mu}=\tensor*{\delta}{^\mu_0}$ is sufficient to effect this, it is not necessary; \emph{depending on the choice of structure constants, one can find non-stationary yet geodesic fluid flows.} Naturally, these correspond to fluid flows representing moving in space (in the chosen frame), so they can represent an implementation of peculiar velocities in the perturbation equations. Such a choice will complicate the equations, though: the directional derivative along the fluid flow does not correspond to a temporal derivative anymore. Much work is done on the influence of peculiar velocities, for instance by Tsagas and collaborators \cite{filippou_large-scale_2021, miliou_peculiar_2024,tsagas_cosmic_2025}, and future research may prove our results a valuable addition.

In a work currently in preparation \cite{scholtens_toymodel}, we build on the results of this study, and in particular the anisotropic perturbation equation \eqref{eq:masterEquations}, by simulating CMBs in general Bianchi universes. More concretely, we will investigate the implementation of this equation in a single scale factor Bianchi V universe, to explore the possible solutions that are admitted. Through the Sachs-Wolfe effect, then, we can relate the so obtained solutions to perturbations as seen in the CMB. Combining this with the non-trivial geometry of the Bianchi V model---in particular the non-trivial null geodesic equations of this geometry---we can make a rough prediction of what a CMB in such a universe would look like. We also offer comparisons to work by Aurich et al.\ \cite{aurich_hyperbolic_2004}, which treated a similar but subtly distinct scenario.

\subsection*{Acknowledgments}
\noindent RWS wishes to thank Sean Gryb for a reading of an initial manuscript, and providing targeted feedback.

%\appendix
%\newpage
%\input{appendices}

\bibliography{library}

\end{document}